\documentclass[amsmath,amssymb,twocolumn]{revtex4}
\usepackage{graphicx, subfigure, amsmath, amssymb, wasysym, chemarr}
\begin{document}

\newcommand{\bgamma}{\mbox{\boldmath $\Gamma$}}
\newcommand{\bL}{\mbox{\boldmath $L$}}
\newcommand{\bT}{\mbox{\boldmath $T$}}
\newcommand{\bI}{\mbox{\boldmath $I$}}
\newcommand{\bM}{\mbox{\boldmath $M$}}
\newcommand{\bN}{\mbox{\boldmath $N$}}
\newcommand{\bE}{\mbox{\boldmath $E$}}
\newcommand{\bA}{\mbox{\boldmath $A$}}
\newcommand{\bB}{\mbox{\boldmath $B$}}
\newcommand{\bD}{\mbox{\boldmath $D$}}
\newcommand{\bG}{\mbox{\boldmath $G$}} 
\newcommand{\bs}{\mbox{\boldmath $s$}}
\newcommand{\bK}{\mbox{\boldmath $K$}}
\newcommand{\bLambda}{\mbox{\boldmath $\Lambda$}}
\newcommand{\bnu}{\mbox{\boldmath $\nu$}}

\newcommand{\bP}{\mbox{\boldmath $P$}}
\newcommand{\bR}{\mbox{\boldmath $R$}}
\newcommand{\bJ}{\mbox{\boldmath $J$}}
\newcommand{\bp}{\mbox{\boldmath $p$}}
\newcommand{\bx}{\mbox{\boldmath $x$}}
\newcommand{\bU}{\mbox{\boldmath $U$}}
\newcommand{\bV}{\mbox{\boldmath $V$}}
\newcommand{\bZero}{\mbox{\boldmath $0$}}
\newcommand{\bLo}{\mbox{\boldmath $L_0$}}
\newcommand{\bNo}{\mbox{\boldmath $N_0$}}
\newcommand{\bNr}{\mbox{\boldmath $N_R$}}
\newcommand{\bSi}{\mbox{\boldmath $S_i$}}
\newcommand{\bSigma}{\mbox{\boldmath $\Sigma$}}
\newcommand{\bSd}{\mbox{\boldmath $S_d$}}
\newcommand{\bdSi}{\mbox{\boldmath $dS_i$}}
\newcommand{\bdSd}{\mbox{\boldmath $dS_d$}}
\newcommand{\bS}{\mbox{\boldmath $S$}}
\newcommand{\bdS}{\mbox{\boldmath $dS$}}
\newcommand{\bds}{\mbox{\boldmath $ds$}}
\newcommand{\bdx}{\mbox{\boldmath $dx$}}
\newcommand{\bsigma}{\mbox{\boldmath $\sigma$}}
\newcommand{\bu}{\mbox{\boldmath $u$}}
\newcommand{\bdt}{\mbox{\boldmath $dt$}}
\newcommand{\bdvds}{\frac{\partial \bv}{\partial \bs}}
\newcommand{\bdvdp}{\frac{\partial \bv}{\partial \bp}}
\newcommand{\bdSdt}{\mbox{$\displaystyle \frac{\bdS}{\bdt}$}}
\newcommand{\bv}{\mbox{\boldmath $v$}}
\newcommand{\el}[2]{\varepsilon^{#1}_{#2}}
\newcommand{\uel}[2]{\widetilde{\varepsilon^{#1}_{#2}}}


\title{
Sensitivity Regulation based on Noise Propagation in Stochastic Reaction Networks
}         
\author{Kyung Hyuk Kim$^1$, Hong Qian$^{2,1}$, and Herbert M. Sauro$^1$\\
 {$^{1}$ Department of Bioenginering, University of Washington, Seattle,
WA 98195}\\
{$^{2}$ Department of Applied Mathematics, University of Washington, Seattle, 
WA 98195}
}
\date{\today}          
\begin{abstract}
In this work we focus on how noise propagates in biochemical reaction networks and affects sensitivities of the system.   We discover that the stochastic fluctuations can enhance sensitivities in one region of the value of control parameters by reducing sensitivities in another region.     Based on this compensation principle, we designed a concentration detector  in which enhanced amplification is achieved by an incoherent feedforward reaction network.   
\end{abstract}

\maketitle

\clearpage

\section{Introduction}

Quantifying biochemical processes at the cellular level is becoming 
increasingly central to modern molecular biology \cite{Elowitz2002, 
Pedraza2005}. Much of this attention can be attributed to the development 
of new methodology for measuring biochemical events in time. In particular, 
the use of green fluorescent protein (GFP) and related fluorophores, high
sensitivity light microscopy and flow cytometry have provided  
researchers with greater details of the dynamics of cellular processes 
at the level of single cells \cite{Elowitz2002,Pedraza2005,Xie2008} 
and even single molecules \cite{Yu2006}.

Quantitative laboratory measurements demand theoretical models 
that are able to describe and interpret the data. The traditional approach
to modeling cellular processes has been to use deterministic equations
with continuous dynamics variables: It assumes that concentrations of
metabolites, proteins, etc. vary deterministically in a continuous
manner. This may be a reasonable assumption when one is dealing with
molecules that occur in relatively large numbers.   For example,
the concentration of ATP in {\it Streptococcus lactis} is approximately
2.5 mM \cite{Maloney1972}   which, assuming a cell volume of
$10^{-15}$ L, yields roughly 1.5 million molecules per cell. In
other cases the number of molecules can be much lower, for example,
the number of {\it LacI} tetrameric repressor proteins in {\it E. coli}
has been estimated to be of the order of 10 to 50 molecules per cell
\cite{Oehler1994}. Such small numbers suggest that a continuum
model may not always be an appropriate description. Moreover,
given the probabilistic nature of chemistry at the molecular level a
deterministic approach appears to be an inadequate description. In
addition recent experimental measurements have clearly
highlighted the stochastic nature of biochemical processes in
individual cells \cite{Elowitz2002}.

Stochastic reaction processes have often been theoretically  investigated using the chemical master equation \cite{Gillespie1992}.  For general reaction systems, this equation is a challenge to solve analytically because the rates of reactions are often expressed as non-linear functions of concentrations \cite{footnote1}.  In addition numerical solutions are equally impractical because of the huge increase in the number of states. Alternative methods such as the Gillespie stochastic simulation algorithm\cite{Gillespie1977},  can also become highly intensive in computation even for reaction systems involving several species of molecules.   Thus, stochastic systems are often modeled with certain approximations for analytical and numerical investigations.    

One such approximation is the linear noise approximation  that has been widely applied to various stochastic systems, to estimate variances and co-variances of concentrations  \cite{Kampen2001}.   The mean levels of concentrations are predicted to be the same as the conventional deterministic approach that neglects stochastic fluctuations of concentrations.   This approximation however becomes invalid as the numbers of molecules in a system get smaller.  Each reaction event becomes more distinguishable and concentrations fluctuate in greater strengths.  The rates of reactions, which are dependent on concentrations,  become more affected by the fluctuations. The mean levels of the rates can become different from the deterministic estimates and this affects the mean levels of concentrations.  The linear noise approximation therefore becomes invalid.   To correct this discrepancy,  different approximation schemes have been introduced: mass fluctuation kinetics (MFK) \cite{Gomez2007} and effective stability analysis (ESA) \cite{Scott2007}.  These approaches provide more accurate analyses at higher noise levels than the linear noise approximation approach,  by taking into account concentration fluctuation effects in mean reaction rates.   MFK uses the moment closure approximation \cite{Kampen2001}  to describe the evolution of the system in terms of the mean and covariance values of concentrations in the course of time.  It has been successfully applied to mass-action type  reaction systems showing stochastic focusing \cite{Paulsson2000} and noise-induced genetic oscillators \cite{Vilar2002}.       It has, however, some limitations on investigating the time evolution of bistable systems because mean and covariance values of the concentrations  cannot distinguish bi-stable and mono-stable systems \cite{Gomez2007}.   For the study of bistable systems especially in stationary states, there exists another approximation approach, ESA.   By using mode-coupling approximations, it successfully describes how concentration fluctuations affect the bistability.   Both MFK and ESA are much less intensive in computation  than the Gillespie's algorithm \cite{Gillespie1977}.

Here we provide an approximate theoretical analysis based on mass fluctuation kinetics  to study stationary state properties of chemical reaction systems.   Our focus in this paper is the investigation of how noise propagation \cite{Paulsson2004, Pedraza2005, Hooshangi2005},  in different network motifs, affects the levels of mean concentrations and mean fluxes.       G\'{o}mez-Uribe et al. \cite{Gomez2007} has shown that the difference between mean rates of reactions and their deterministic (without any noise) estimates can be predicted from the concentration covariances and the curvatures of the nonlinear rate functions (reaction rate equations), as a first order contribution \cite{Gomez2007}.   We have investigated this \emph{curvature-covariance effect}  more thoroughly and have found that the effect provides simple and clear qualitative illustrations  of stochastic focusing, and leads to qualitative understanding on how to design and control chemical reaction systems to achieve certain noise-responses in system behaviors.

The curvature-covariance effect shows that increased sensitivity (stochastic focusing) in one region of the values of control parameters can lead to decreased sensitivity (stochastic defocusing) in other regions.  Here the sensitivity is a  measure of a system response due to a source signal change and it is  defined as the ratio of the percentage change of a response signal ($r$)  to the percentage change of a source signal ($s$):
\[
\mbox{Sensitivity} = \frac{s}{r}\frac{d r}{d s} = \frac{d \ln r}{ d \ln s}.
\]
  We have applied this \emph{stochastic focusing-defocusing compensation effect} to investigate an incoherent feedforward concentration detector \cite{Mangan2003, Entus2007}.      The concentration detector shows increased amplification of the concentration detection.  This is due to the   stochastic focusing.  The sensitivity of the detection is however not enhanced due to  the stochastic defocusing.  By tuning system parameters, we could enhance  the amplification up to eight times.  By applying the curvature-covariance effect, we have further increased the amplification, by modifying upstream subnetwork structures.

We present our analysis in this manuscript as follows.   In section~\ref{sec:sca} we will show  how both concentration fluctuations and the curvature of the rate functions  affect the mean rates of reactions.  We also explain the mechanism of the stochastic focusing-defocusing compensation.   In section \ref{sec:application} we will  illustrate the fluctuation effects in  various network motifs, which will also be tested by Monte-Carlo simulations by using the Gillespie stochastic simulation algorithm.   We will investigate negative feedback (homeostasis, hyperbolic inhibition) and incoherent feedforward (concentration detection) regulation.   

Our analysis, like other approximation approaches introduced earlier,  is based on the chemical master equation and thus the noise is considered intrinsic \cite{Kampen2001}, which means the noise is generated by random chemical reactions involved in reaction systems and all other noise from outside the system is considered negligible.     As in MFK and ESA, our analysis becomes invalid  if  the third and higher moments of the noise correlation need to be taken into account \cite{Gomez2007, Scott2007}.     Our analysis focuses only on the stationary state behaviors of the processes without any oscillation, which can be further investigated as an extension of this proposed analysis.

\section{Results}\label{sec:sca}
\subsection{Curvature-Covariance Contribution to Mean Reaction Rates}
We consider chemical reaction networks where the number of particles involved in the reactions are low.  Due to the small particle numbers, the change in the number   of particles due to chemical reactions is observed as a discrete process \cite{Kampen2001}.    In addition, the reactions occur due to the random collisions between reactants.  Thus, the change of molecule numbers is not only discrete but also random and is often described by discrete stochastic processes.   Under special conditions (homogeneous and statistically independent reaction events), stochastic processes are  fully described by the chemical master equation \cite{Gillespie1992}.  This equation describes the time evolution of a probability distribution function, which represents the probability that one finds the number of molecules for each species at a given time.  The stochastic reaction events cause molecule concentrations to fluctuate and this determines the rates of the next reaction events.  These events again cause the concentrations to fluctuate and the same argument is applied to the following steps.  The mean values of the rates can be affected by the concentration fluctuations.   This effect will be formulated in this section.

We will characterize the concentration fluctuations by using mean values and co-variances of the concentrations: E.g., in the fluorescent protein experiments,  fluctuations in light intensity can be measured in time.  Once the light intensity fluctuation stabilizes at  a constant level (in the stationary state), this level indicates the relative mean value of the concentration of the light emitting proteins, and the standard deviation from the constant level measures the relative strength of the fluctuations.  If the experiments are performed by using both green and yellow fluorescent proteins, one can measure the intensities of both proteins and quantify how much the protein concentrations fluctuate together (fluctuation correlation)  by measuring their co-variance.  

We consider $m$ species of molecules and $n$ reactions.  For our purpose we assume that system parameters $\bp$ are
non-fluctuating (bold symbols represent matrices and vectors).  $\bp$ can be reaction rate constants in mass action rate equations, dissociation constants in Hill equations, temperature, pH, etc.  The time evolution  of mean concentrations and concentration co-variances can be described by the following equations \cite{Gomez2007}(see Appendix~\ref{appendix-mean-variance}):
\begin{eqnarray}
\frac{d\langle \bs \rangle }{dt} &=& \bN_R  \Bigg(\bv( \langle \bs \rangle, \bp  ) + \sum_{ij}\frac{1}{2}\frac{\partial^2 \bv(\langle \bs \rangle, \bp ) }{\partial \langle s_i \rangle \partial \langle s_j \rangle}  \sigma_{ij}\Bigg),
\label{mean1}
\end{eqnarray}
\begin{eqnarray}
\frac{d\bsigma}{dt} &=&  \Big[ \bJ \bsigma+  \bsigma^T  \bJ^T + \frac{ \bD(\langle \bs \rangle ) }{\Omega}\Big],  \label{variance1} 
\end{eqnarray}
where $\bN_R$ is a $m_0 \times n$ reduced stoichiometry matrix \cite{Ingalls2003}.  $m_0$ is the number of linearly independent rows in a stoichiometry matrix.   $\bv$ represents \emph{propensity functions} describing the rates of reactions: $\bv \equiv \{ v_1, \cdots, v_n\}$.    We assume that $v_i$ is a function of both $\{s_j\}$ (with $j=0, \cdots, m_0$) and $p_i$, i.e,  $v_i(\bs, p_i)$.  The angle brackets denote the average: more precisely, the average of data taken at a given time $t$ for repeated independent runs of simulations or experiments.  $J_{ij}$ is an element of the Jacobian matrix defined as
\[
J_{ij}\equiv \sum_{k=1}^n N_{R_{ik}}  \frac{\partial v_k(\langle \bs \rangle, p_k )}{\partial \langle s_j \rangle}.
\]
The covariance matrix $\bsigma$ is defined as
\[
\sigma_{ij} \equiv \Big\langle (s_i -\langle s_i \rangle )(s_j -\langle s_j \rangle )\Big\rangle.
\]
For $i=j$, the covariance becomes the variance.  The matrix $\bD$ is the diffusion coefficient matrix defined by $ \bN_{R} \bLambda \bN_{R}^T $ with a diagonal matrix $\Lambda_{ij} \equiv v_i\delta_{ij}$.  $\Omega$ is the system volume.  The above equations do not hold exactly but are valid under certain approximations (see Appendix~\ref{appendix-mean-variance} and \ref{appendix-truncation}).  The validity of the above equations, however,  should be checked on the basis of  each different model since noise in adjacent systems can be propagated into the propensity function \cite{Paulsson2004, Pedraza2005, Hooshangi2005}.  We will show several examples of this noise propagation in Example~2.   We note that Eq.~\ref{variance1} is different from the results of G\'{o}mez-Uribe, et al.~\cite{Gomez2007}; We have neglected all the terms of the order of $1/\Omega^2$ that have been kept in Eq.~11 of G\'{o}mez-Uribe, et al. \cite{Gomez2007}.  For a detailed discussion readers are referred to Appendix \ref{appendix-mean-variance} and  \ref{appendix-truncation}.  This is consistent within the approximation of  truncation of third and higher order moments.

\begin{figure}[h!]
  \begin{center}
\includegraphics[scale=0.5, angle=0]{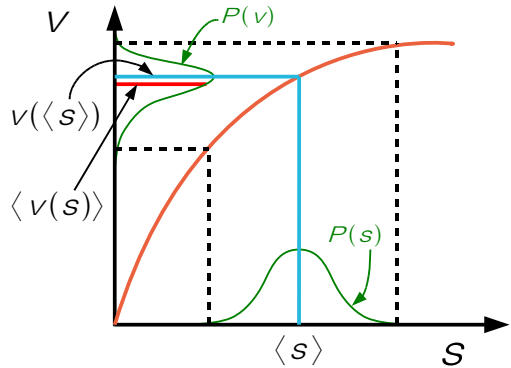}
   \end{center}
 \caption{A mean value of a Michaelis-Menten propensity function $v(s)$ depends on its curvature and the variance of substrate concentration $s$.  The probability distribution function of $s$, $P(s)$, is assumed to be symmetric.  The probability distribution function of $v(s)$, $P(v)$, becomes asymmetric due to the nonlinear propensity function $v(s)$.  This makes the difference between $v(\langle s \rangle)$ and $\langle v(s) \rangle$.}
  \label{fig:cur-var-effect}
\end{figure}

The right-hand term in parentheses in Eq.~\ref{mean1} is called the \emph{mean propensity function} and shows that the mean rate of reaction, denoted by $\bnu$, is affected by concentration covariances $\bsigma$ and the curvature of the propensity function ($\bv$), $\partial^2 \bv/\partial s_i \partial s_j$.    To illustrate this \emph{curvature-covariance correction}, let us consider an example: a substrate is converted to a product through an enzyme reaction and the enzyme-substrate complex turnovers so fast that we can assume that the complex is in quasi-steady state.  Then, the creation rate of the product can be described by the Michaelis-Menten rate equation $v(s)$ \cite{Haseltine2002, Rao2003, Goutsias2005}:
\[
\longrightarrow S \xrightarrow{v(s)}P.
\]
We assume that fluctuations in  the concentration of molecule $S$, $s$, are symmetric with respect to its mean value.   Then, the fluctuation of $v$ becomes asymmetric because the positive fluctuations of $s$ cause the fluctuation of $v$ to become relatively smaller than the negative fluctuations of $s$,  as shown in Fig.~\ref{fig:cur-var-effect},  due to the curved shape of the propensity function $v(s)$.     As the curvature of the propensity function increases, the probability distribution of $v$ becomes more asymmetric and the mean value of the propensity function $\langle v (s) \rangle$ deviates  from the deterministic prediction $v (\langle s \rangle)$.    As the distribution function of $s$ gets narrower, the distribution function of $v$ also gets narrower and the difference between $v(\langle s \rangle)$ and $\langle v(s) \rangle$ gets smaller.   This is why Eq.~\ref{mean1} shows that the first order of correction to a deterministic rate is proportional to both  concentration covariances and the curvature of a propensity function.

The non-zero covariances occur in general because the concentration fluctuation in one species is propagated into that of another species via the connected paths of reaction networks and such propagated  fluctuations cause the propensity function to fluctuate.   Such non-local effects are incorporated into the covariance term in Eq.~\ref{mean1} and we will give an example of this effect in section~\ref{sec:application} Example 2.

We now focus on stationary state properties of Eq.~\ref{mean1} and \ref{variance1}.  In the stationary state, the right hand sides of these equations vanish.  Eq.~\ref{variance1} leads to a Lyapunov equation showing how concentration covariances $\bsigma$  are related to $\bD$ (strengths of random driving forces causing the concentrations to fluctuate) and $\bJ$ (strengths of tendencies returning to stable fixed points of the case in the absence of the random driving forces):
\begin{equation}
 \bJ  \bsigma+ \bsigma^T  \bJ^T + \frac{ \bD(\langle \bs \rangle ) }{\Omega}=0. 
\label{variance2}
\end{equation}
The above equation can be considered as a non-equilibrium steady-state extension of  the equilibrium fluctuation-dissipation theorem \cite{Kubo1966, Paulsson2004}.   By solving the above equation for $\bsigma$, $\bsigma$ can  be expressed as a function of $\langle \bs \rangle$.  We denote the solution by $\bsigma^*(\langle \bs \rangle, \bp)$. We substitute this into Eq.~\ref{mean1}:
\begin{equation}
\bN_R  \bnu(\langle \bs \rangle, \bp)=0, 
\label{mean2}
\end{equation}
where 
\begin{equation}
\bnu (\langle \bs\rangle,  \bp) \equiv \bv(\langle \bs\rangle, \bp) + \sum_{ij} \frac{1}{2}
   \frac{\partial^2 \bv}{\partial s_i\partial s_j}\Big|_{\bs=\langle \bs\rangle}
\sigma^*_{ij}(\langle \bs \rangle, \bp).
\label{eqn:vtilde}
\end{equation}
We have denoted the mean propensity function using $\bnu$.  The mean propensity function $\bnu$ becomes a true measurable quantity when Eq.~\ref{mean2} is solved for $\langle s \rangle$ and its solution, denoted by $\bs^*$, is substituted in Eq.~\ref{eqn:vtilde}.

This shows that the traditional deterministic propensity function must be corrected in order to yield the `true' measurable mean rate in the stochastic regime by  replacing $\bv( \bs^*, \bp)$ by $\bnu(\bs^*  , \bp)$.  This implies that all the theorems of metabolic control analysis \cite{Kacser1973, Fell1992, Fell1996, Ingalls2006}, which describes how the changes in enzyme activities affect metabolite concentration levels,  can be applicable to the stochastic reaction systems (for more discussion on this, we refer to section \ref{sec:conclusion}).

\subsection{Stochastic Focusing-Defocusing Compensation}
We investigate the properties of the mean propensity function $\bnu$: Eq.~\ref{mean2}.  The curvature-covariance effect leads to an intuitive understanding of the origin of stochastic focusing and furthermore it allows to discover a typical feature of the stochastic focusing that the stochastic fluctuations can enhance sensitivities in one region of the value of control parameters by reducing sensitivities in another region.  We call this \emph{stochastic focusing-defocusing compensation}.  This compensation typically appears in inhibition regulation and sigmoidal responses.  

Stochastic (de-)focusing is a phenomenon where system sensitivities are enhanced (reduced) due to stochastic fluctuations \cite{Paulsson2000}.   To understand the stochastic focusing-defocusing compensation effect, we consider the following reaction system:
\begin{equation}
\xrightarrow{c_0} S \xrightarrow{c_1 s} \o,~~\xrightarrow{v(s)}P \xrightarrow{k p} \o,
\label{eqn:SDF}
\end{equation}
where $s$ and  $p$ are the concentrations of $S$ and $P$ respectively and $c_1$ and $k$   degradation rate constants of $S$ and $P$ respectively. $c_0$ is the creation rate of $S$.    We observe how a response signal (chosen to be $\langle p \rangle$) changes due to the perturbation of a source signal ($\langle s \rangle $).    The sensitivity is defined as 
\[
\mbox{Sensitivity} = \frac{d \ln \langle p \rangle }{d \ln \langle s \rangle} =\frac{d \ln (\nu(s^*)/k) }{ d \ln s^*} = \frac{d \ln \nu(s^*) }{ d \ln s^*},
\]
where we have used the fact that in the stationary state the mean concentration of $p$  is equal to $\langle v(s) \rangle/k \simeq \nu(s^*)/k$ because the mean creation rate of $P$ balances its mean degradation rate.   Thus,  we investigate how $\nu$ changes due to stochastic fluctuations. 

\begin{figure}[h!]
  \begin{center}
\includegraphics[scale=0.6, angle=0]{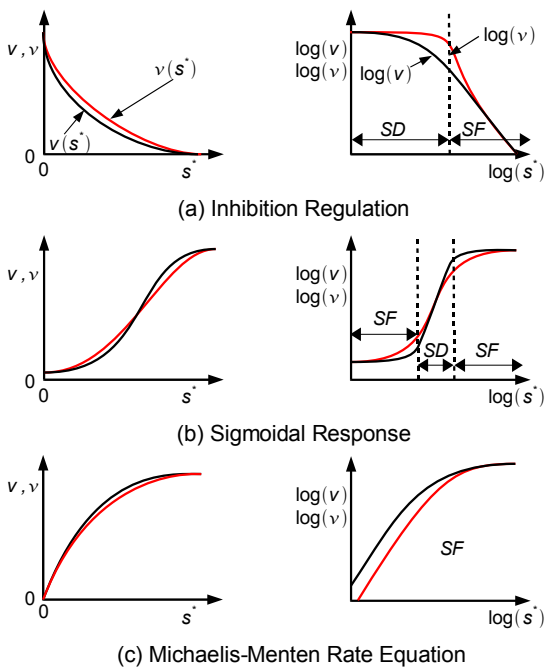}
   \end{center}
 \caption{Stochastic focusing-defocusing compensation: Three different types of propensity functions for a reaction scheme Eq.~(\ref{eqn:SDF}) show different compensation patterns.  $s^*$ is a mean level estimate of concentration $s$, which can be varied by changing a system parameter such as the creation rate of $s$.   Depending on this parameter value, stochastic focusing (SF) or stochastic defocusing (SD) appears.   Inhibition regulation and sigmoidal response of $v(s)$  lead to the  compensation [(a) and (b)].  Michaelis-Menten type rate equation can result in only SF without SD as shown in (c). 
Solid black lines corresponds to $v(s^*)$ and red lines to $\nu(s^*)$. 
}  
  \label{fig:SDF}
\end{figure}

We consider a sigmoidal response in $v(s)$ given by $k_1 + \frac{k_2 s^3}{k_3 + s^3}$ with $k_i$ positive constants for $i=1,2,3$.   
As shown in Fig.~\ref{fig:SDF}(b), the curvature of $v(s)$ changes from positive sign to negative as $s$ increases from zero.    The sign of $\nu-v$  changes from positive to negative due to the curvature-covariance effect, and $\nu-v$ converges to zero as $\langle s \rangle  \rightarrow \infty$ and $\langle s \rangle  \rightarrow 0$ since the curvature vanishes in these limits.   Therefore, stochastic defocusing (SD) appears in between two stochastic focusing (SF) regions as shown in Fig.~\ref{fig:SDF}(b).  

If $v(s)$ represents a hyperbolic-type  inhibition of $P$ by $S$: $v(s) = k_1/(k_2+s)$, the curvature of $v$ is positive for all $s$ except $s\rightarrow \infty$.  The variance of $s$ vanishes when $\langle s \rangle \rightarrow 0$.   Thus, $\nu-v$ is always positive except that $\langle s \rangle =0$ and $\infty$.  This means that SD changes to SF as $\langle s \rangle $ increases from zero as shown in Fig.~\ref{fig:SDF}(a).  

However, SF does not always come with SD.   For example, when $v(s)$ is a Michaelis-Menten type propensity function, only SF appears without SD as shown in Fig.~\ref{fig:SDF}(c).

We hope that this \emph{stochastic focusing-defocusing compensation effect}  and \emph{the curvature-covariance effect} can be applied to design and control reaction networks to improve system functionality by exploiting intrinsic noise \cite{Mangan2003, Entus2007}.   As an application, in section \ref{sec:application}, we will design incoherent feedforward concentration detectors to improve detection amplification by applying our results.

\section{Applications}
\label{sec:application}
\subsection*{Example 1: Curvature-covariance Contributions to Mean Propensity Functions}
To examine  in detail the contribution that the curvature-covariance makes to the mean propensity function,  we will present three examples.  In the first example, we consider an association reaction \cite{Gomez2007}:
\[
S_1+ S_2 \xrightarrow{v=k s_1 s_2} X. 
\]
In the stationary state, the association propensity function $v$ has its mean value at 
\[
\langle v \rangle  = k \langle s_1 s_2 \rangle = k \langle s_1 \rangle \langle s_2 \rangle + k \langle (s_1- \langle s_1 \rangle )(s_2 -\langle s_2\rangle ) \rangle. 
\]
The first term on the right hand side is the propensity function for deterministic systems. 
The second term  is the covariance between $s_1$ and $s_2$. Thus,  the true mean value of the reaction rate can be estimated by taking into account the covariance effect. 

As a second example, we consider a Michaelis-Menten-type reaction:
\begin{equation}
\varnothing \xrightarrow{v_0} S \xrightarrow{v(s)} X,
\label{eqn:MM}
\end{equation}
where $v_0$ is a fixed creation rate of $S$ and $v(s) \equiv \frac{V_{max} s}{ K_m + s}$,
with $K_m$ the Michaelis-Menten constant  and  $V_{max}$ a saturation rate.  If the variable  $s$ is non-stochastic, then the probability distribution function of $s$ is a delta-function centered at $\langle s \rangle$.  Thus, $\langle v(s) \rangle  = v(\langle s \rangle)$.  However, if $s$ fluctuates stochastically, this equality does not hold any more.  The fluctuations in $s$ cause fluctuations in the propensity function.  Since the propensity is a curved-down function in $s$, the negative direction of fluctuation in $s$ will cause the fluctuation in $v$ to be more negative than the positive direction of fluctuation in $s$.  Such biased fluctuations in $v$ makes $\langle v (s) \rangle$ smaller than $v( \langle s \rangle )$.    For small enough fluctuations such differences can be shown to be proportional to the curvature of $v$ and also to  the concentration variance in $s$:
\begin{equation}
\nu(s^*) = v(s^*) + \frac{1}{2}\frac{\partial^2 v}{\partial s^2}\Big|_{s=s^*} \sigma^*.
\label{eqn:mean-rate}
\end{equation}
The above equation is derived from Eq.~\ref{eqn:vtilde}.  $s^*$ is the solution of Eq.~\ref{mean2} and $\sigma^*$ the solution of Eq.~\ref{variance2} in the stationary state.
Such curvature-variance correction to the deterministic propensity function is closely related to stochastic focusing \cite{Paulsson2000} (see Example~3 and 4). 

\begin{figure}[h!]
  \begin{center}
\subfigure[Propensity Functions for $K_M=5$]{\includegraphics[scale=0.6, angle=-90]{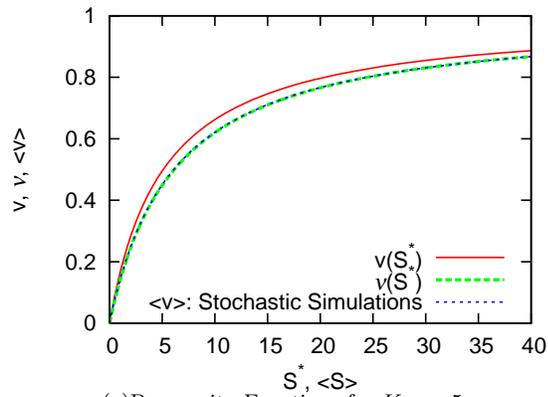}}
\subfigure[Probability Distribution Functions] {\includegraphics[scale=0.6, angle=-90]{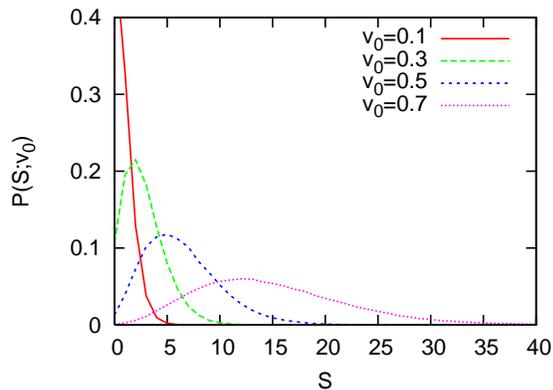}}
   \end{center}
 \caption{Molecules $S$ are created with a constant rate $v_0$ and degrades with a Michaelis-Menten  rate $v(S)=S/(K_M+S)$ as shown in Eq.~\ref{eqn:MM}.  (a) The stationary state degradation rate in both the stochastic and deterministic cases are compared with numerical simulations for fixed value of  $K_M=5$ for various creation rates.  The Gillespie stochastic simulation algorithm is used.  $\langle v(S) \rangle$ and $\langle S \rangle$ are estimated by time averages of a single run of simulation.  $S^*$ denotes the approximate estiamate of $\langle S \rangle$  given by solving Eq.~\ref{mean2}.  (b) Probability distributions in $S$ are shown for each different value of $v_0$.  }  
  \label{fig:MM-obc}
\end{figure}

\begin{figure}[h!]
  \begin{center}
\subfigure[$K_M=1$]{\includegraphics[scale=0.6, angle=-90]{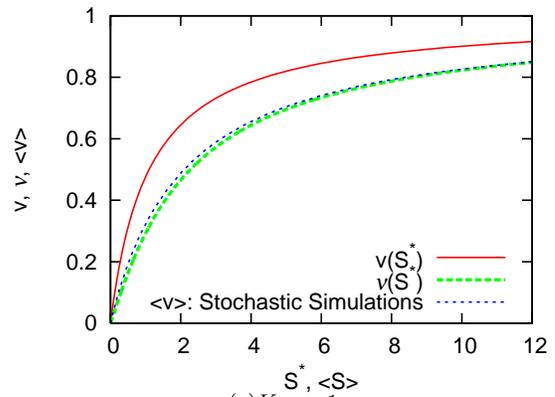}}
\subfigure[$K_M=10$]{\includegraphics[scale=0.6, angle=-90]{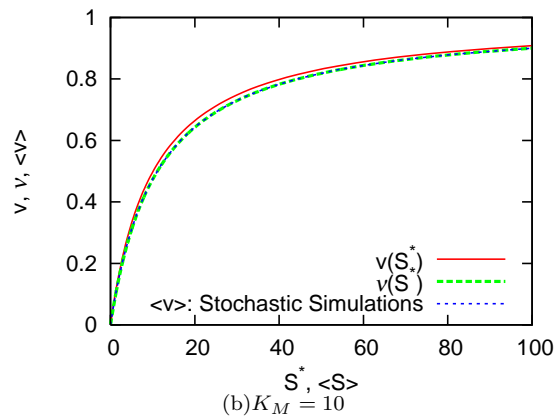}}
   \end{center}
 \caption{Degradation propensity functions in a reaction system Eq.~\ref{eqn:MM} for different values of $K_M$.}
  \label{fig:MM-obc2}
\end{figure}

Monte-Carlo simulations based on the Gillespie stochastic simulation algorithm \cite{Gillespie1977} were performed for the above Michaelis-Menten-type reaction.  We set $V\equiv S/(K_M + S)$ where we rescale time so that the maximum rate of $V$ is set to one and varies $V_0$ from $0$ to $0.9$ for given values of $K_M$.   An upper case letter $S$ denotes the molecule number of species $S$ and the lower case letter $s$ its concentration, i.e., $s=S/\Omega$.  The propensity function $V$ also needs to be divided by the system volume $\Omega$ to become the propensity function $v$ defined in the previous section.  To avoid too many notations, we take the same symbol for both the propensity functions in spite of the difference.  $K_M$ needs to be also divided by $\Omega$ to become the Michaelis-Menten constant.  For $K_M \gtrsim 1$, the variance corrections become accurate  as shown in Fig.~\ref{fig:MM-obc} and \ref{fig:MM-obc2}.     For $K_M=5$ (Fig.~\ref{fig:MM-obc}(b)), when $v_0 \simeq 0.1$, $S$ fluctuates between $0$ and $4$ for most of the time in the stationary state, where the propensity function is almost linear in $S$ (Fig.~\ref{fig:MM-obc}(a)). Thus, the distribution of $S$ becomes similar to the Poisson distribution.  This is why the variance correction becomes negligible for this range of the value of $S$.  For $v_0 \simeq 0.9$ the reaction is saturated and the propensity function becomes almost linear again.  The variance correction becomes negligible.  For $v_0$ values between $0.4$ and $0.7$ the variance correction becomes significant but still gives reasonable  estimates for the mean propensity.  For the smaller values of $K_M$, the corrections become less accurate.  This is because $\partial^3 v / \partial s^3$ becomes larger and the  neglected terms in Eq.~\ref{eqn:mean-rate} become significant.   Here, we have shown the origin of the curvature-covariance effect on the mean propensity functions and have also illustrated how accurate the moment closure approximation is.   The approximation needs to be verified however case by case;  e.g., the creation rate $v_0$ in the above example was assumed to be constant but it can be a function of other species concentrations, noise of which, possibly significant amounts of noise,  can be propagated into $v_0$.

We are going to investigate reaction systems with non-linear propensity functions, e.g., Michaelis-Menten kinetics.  This is because various network motifs can be directly represented by propensity functions.  Such non-linear rate equations  can be realized  in mass-action reaction systems, when  the time scales of reactions can be separable into slow and fast ones and the molecules involved in the fast reactions are treated as in stationary states \cite{Haseltine2002, Rao2003, Goutsias2005}.    This quasi-steady state approximation leads to non-linear rate equations with \emph{simplified} network structures.  However, this approximation results in neglecting  all the correlations between the fast and slow concentration fluctuations.  If the time scales are not separable, such correlations become significant and the results derived by using the nonlinear propensity functions can be significantly different from the true measured ones.  It is however important to investigate the simplified network by using the nonlinear propensity functions as a first step toward to understand the system behavior related to each different network motif. 

\begin{figure}[h!]
\begin{center}
\subfigure[A Detailed Reaction Network]{\includegraphics[scale=0.4]{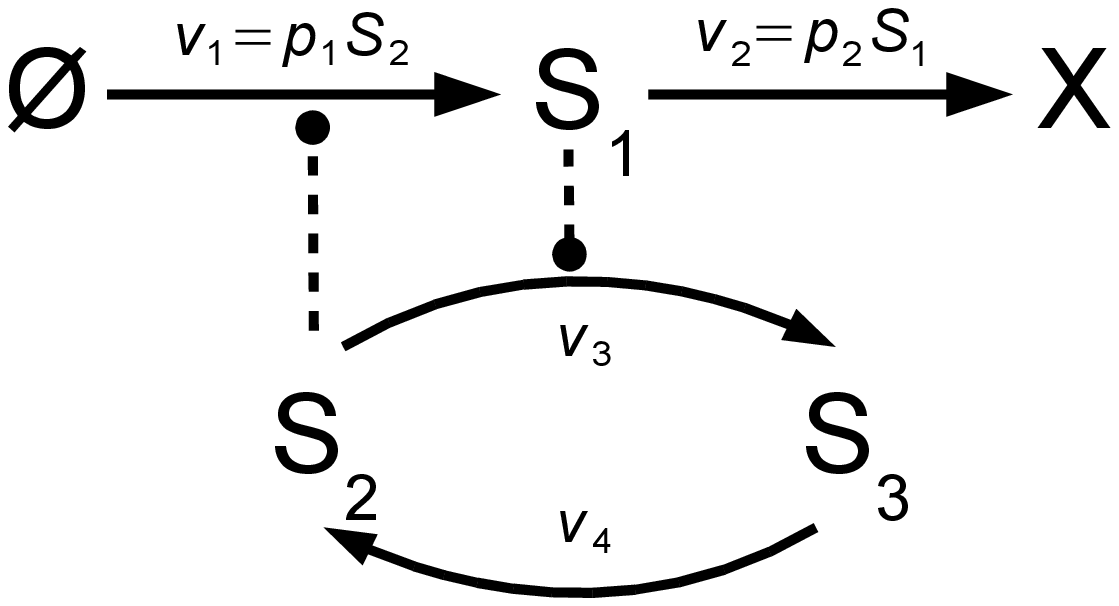} \label{fig:nfb-cycle}}
\subfigure[A Simplified Reaction Network]{\includegraphics[scale=0.4]{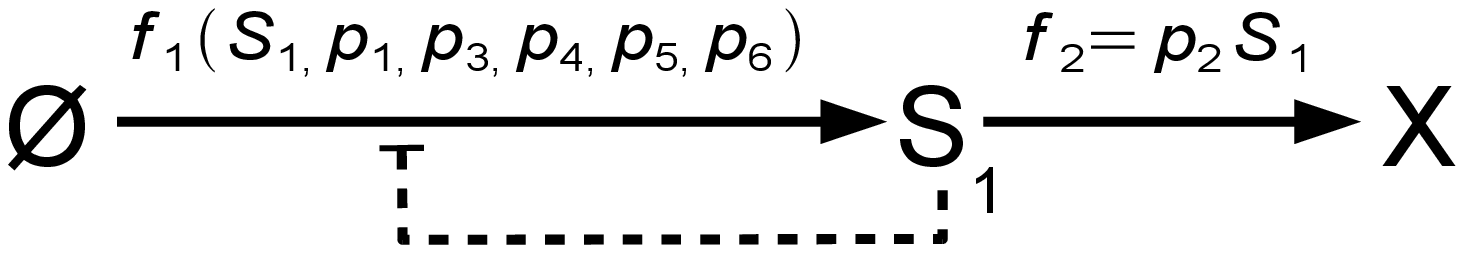}\label{fig:nfb-cycle-simplified}}
\end{center}
  \caption{(a) Cycle reactions involving $S_2$ and $S_3$ act as a negative feedback on $S_1$.  The rates of the cycle reactions are given as $v_3=p_3 S_1 S_2/(p_5+S_2)$ and $v_4=p_4 S_3/(p_6+S_3)$ \cite{Tyson2003}. (b) The reaction network shown in (a) becomes simplified when the cycle reactions are fast.   $f_1$ represents a negative-feedback propensity function (for its detail functional form, we refer to Tyson et al.~\cite{Tyson2003}).  A circular (flat) end dotted line corresponds to positive (negative) regulation. }
 \end{figure}

To illustrate how reasonable it is to use the nonlinear propensity functions, we consider a negative feedback system showing homeostasis: A protein species $S_1$ enhances the phosphorylation of another protein species and its dephosphorylated form ($S_2$) inhibits the generation of protein $S_1$ as shown in Fig.~\ref{fig:nfb-cycle}.   
We assume that the phosphorylation-dephosphorylation cycle ($v_3$ and $v_4$) is very fast so that the fluctuation of $S_1$ immediately appears in the concentration of $S_2$ and inhibits the generation of $S_1$.  We also assume that the number of molecules involved in cyclic reactions are large enough that the internal noise due to the phosphorylation-dephosphorylation is negligible.   In this case, the above reaction can be further simplified to a negative feedback reaction system as shown in Fig.~\ref{fig:nfb-cycle-simplified}.  

\begin{figure}[h!]
  \begin{center}
 \subfigure[Propensity Functions]{\includegraphics[scale=0.5, angle=-90]{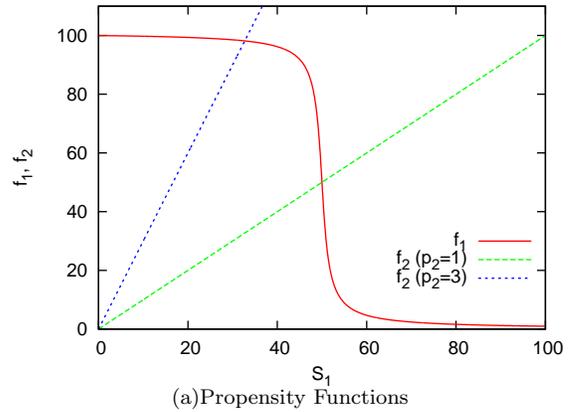}}
{
\subfigure[$P(S_1)$]{\includegraphics[scale=0.5, angle=-90]{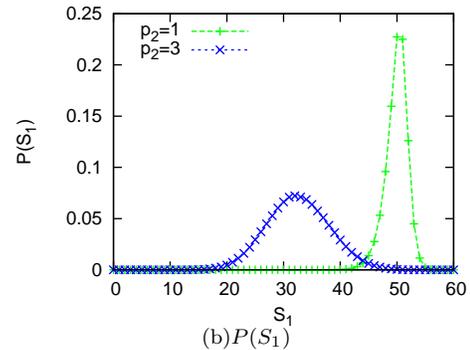}\label{fig:homeo-b}}

}
  \end{center}
  \caption{Negative feedback homeostasis: The number fluctuation is suppressed by a negative feedback as shown in Fig.~\ref{fig:nfb-cycle-simplified}.  The fluctuation strength decreases as the mean number of $S_1$ increases by changing $p_2$ from $3$ to $1$.  (See Tyson, et al.~\cite{Tyson2003} for the relationship between the detailed one and the simplified one: $p_1=1$, $p_5=1$, $p_6=1$ and $S_2+S_3=100$.)}
  \label{fig:homeo}
\end{figure}

This system can show homeostasis due to a strong negative feedback.   In Fig.~\ref{fig:homeo}, the negative feedback becomes very strong within a narrow range of $S_1$ centered around $S_1=50$; the concentration of $S_1$ can be decreased right after the degradation is accelerated by increasing $p_2$ but the concentration eventually return  to the approximate value of the original concentration because the creation rate is strongly increased.  This phenomenon is called homeostasis.   Homestasis has been shown to be related to suppressing concentration fluctuations (see Fig.~\ref{fig:homeo-b} and \cite{Elf2003}).  The strong negative feedback corresponds to the large difference between unscaled elasticities \cite{Fell1992} of reactions $v_1$ and $v_2$.   Such a  large difference means that when the concentration fluctuates with respect to the mean value of the concentration, the system has a strong tendency to dampen  the fluctuation and return to the mean value.  When the value of $S_1$ is within the range, where homeostasis appears ($p_2=1$ in Fig.~\ref{fig:homeo-b}), the probability distribution of $S_1$ becomes narrow.  When $p_2$ increases to $3$, however, the homeostasis vanishes and the probability distribution of $S_1$ shows very large fluctuations.  

\begin{figure}[h!]
\begin{center}
{\includegraphics[scale=0.5, angle=-90]{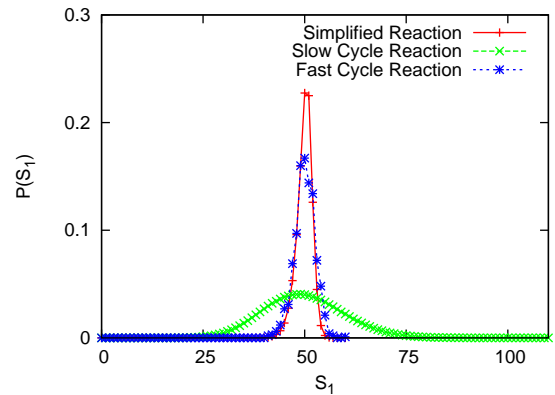}}
\end{center}
  \caption{The probability distribution of $S_1$ for the non-simplified reaction Fig.~\ref{fig:nfb-cycle} and simplified one Fig.~\ref{fig:nfb-cycle-simplified}. For both the slow and fast cycle reactions, we have used the following parameters $p_1=1$, $p_2=1$,  $p_5=1$, $p_6 = 1$ and $S_2+S_3=100$.  For the slow reaction, $p_3=0.01$ and $p_4=0.5$ are used.  For the fast reaction,  $p_3=1000$ and $p_4=50000$. }
  \label{fig:homeo-compare}
\end{figure}

We compare the above two non-simplified and simplified reaction systems, Fig.~\ref{fig:nfb-cycle} and \ref{fig:nfb-cycle-simplified}. As the cyclic reaction gets faster, the distribution function  of $S_1$ for the non-simplified reaction converges to that for the simplified one as shown in Fig.~\ref{fig:homeo-compare}.    This shows how reasonable the use of non-linear propensity functions is.   However, if the cyclic reactions are not fast, then the correlation between the fast and slow variables $S_1$ and $S_2$ ($S_3$) become significant and the system behavior becomes very different from the one with nonlinear propensity functions.  The variance of $S_1$ can increase rather than decrease as $p_2$ changes from 3 to 1 for the slow reaction case (The graph for this result not shown).

\subsection*{Example 2: Non-local Mean Propensity Functions}
\begin{figure}[h!]
\begin{center}
\subfigure[A Detailed Reaction Network]{\includegraphics[scale=0.4]{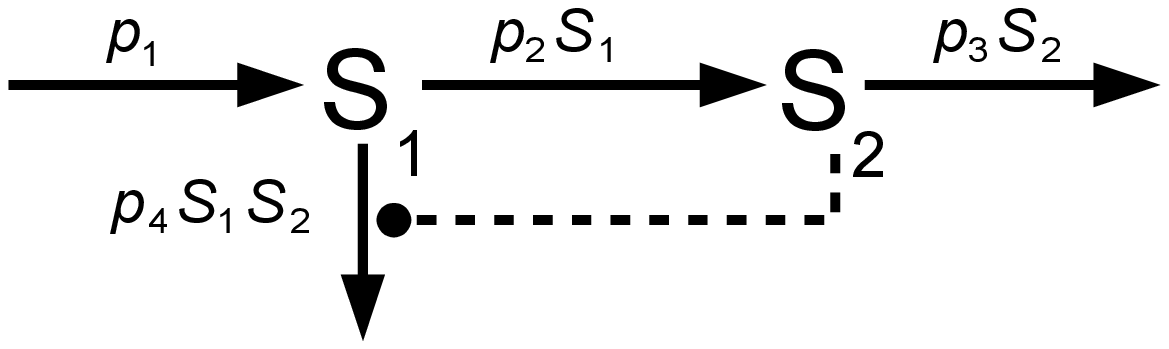} \label{fig:nfb-hyper}}
\subfigure[A Simplified Reaction Network]{\includegraphics[scale=0.4]{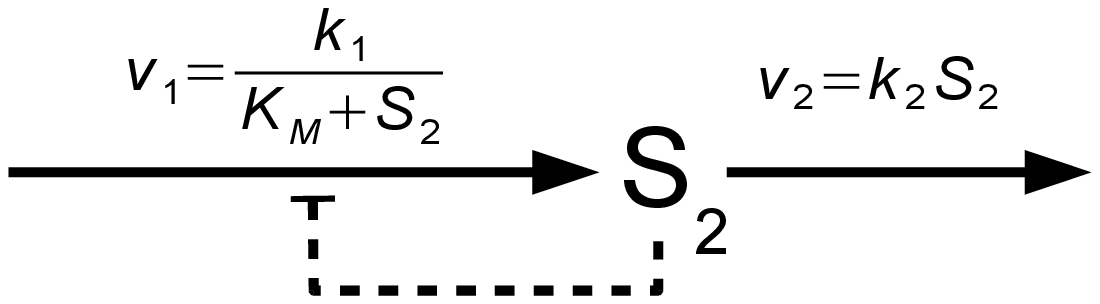}\label{fig:nfb-hyper-simplified}}
\end{center}
  \caption{A hyperbolic inhibition reaction network:  In (a),  a product of $S_1$ accelerates the degradation of itself.   In (b), this reaction gets simplified into a hyperbolic-type inhibition reaction when the reactions involving $S_1$ is much faster than that of $S_2$.  $k_1 = p_1 p_2/p_4$, $K_M = p_2/p_4$, and  $k_2 = p_3$.}
 \end{figure}

As shown in Example 1, the mean propensity function is affected by the noise covariance $\bsigma$.  Here we discuss the non-local property in the noise covariance.   We consider a negative feedback reaction system as shown in Fig.~\ref{fig:nfb-hyper}.  $S_1$ produces $S_2$, which accelerates the degradation of $S_1$. This reaction system can be further simplified as in Fig.~\ref{fig:nfb-hyper-simplified} when the life time of $S_1$ is much shorter than that of $S_2$.    The propensity function $v_1$ does not depend on $k_2$, but its mean value $\nu_1$ becomes dependent on $k_2$ in the stationary state because the concentration variance depends on $k_2$; the concentration fluctuations are due to the events of both the reactions, so the concentration variance of $s$ depends  not only on $v_1$ but also on $v_2$:
\begin{eqnarray*}
\lefteqn{\nu_1(\langle s \rangle, k_1, k_2)} \\
&=& \frac{k_1}{K_M+\langle s \rangle}+ \frac{k_1(k_1+k_2 \langle s \rangle (K_M+\langle s \rangle))}{2(K_M+\langle s \rangle)^2 (k_1+k_2(K_M+\langle s \rangle)^2)},
\end{eqnarray*}
where we have expressed the variance $\sigma$ in terms of $\langle s \rangle$ by using  Eq.~\ref{variance2} and \ref{eqn:mean-rate}.

\begin{figure}[h!]
\begin{center}
{\includegraphics[scale=0.4, angle=0]{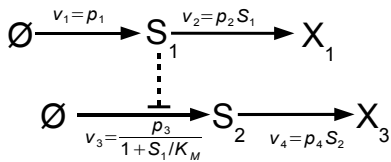}}
\end{center}
  \caption{Stochastic focusing and defocusing by using a two-step cascade reaction system, where $S_1$ inhibits the creation of $S_2$. }
  \label{fig:sf}
\end{figure}

As a second example, we consider a two-step cascade reaction system studied by Paulsson et al. \cite{Paulsson2000} as shown in Fig.~\ref{fig:sf}.  Signal molecules $S_1$ inhibits the production of $S_2$.  The fluctuation in $S_1$ is propagated into the propensity function $v_3$.  Thus, the mean propensity function $\nu_3$ will depend not only on $p_3$ and $p_4$ but also on $p_1$ and $p_2$ due to noise propagation from an upstream reactions $v_1$ and $v_2$.   We will study this system in detail in the next section.

\subsection*{Example 3: Stochastic Focusing-Defocusing Compensation}
The curvature-covariance effect can explain stochastic focusing, which describes the phenomenon that system sensitivities are enhanced due to stochastic fluctuations \cite{Paulsson2000}.

We consider again the example of the cascade reaction as shown in Fig.~\ref{fig:sf}.     In Paulsson et al.~\cite{Paulsson2000}, stochastic focusing was studied by perturbing the parameter $p_1$ and examining how sensitive the mean concentration of $S_2$ is  to the change in the mean concentration of $S_1$.  We can quantify such sensitivity in terms of concentration control coefficients:
\begin{equation}
C_{p_1} = \frac{C^{\langle S_2 \rangle }_{p_1}}{C^{\langle S_1 \rangle}_{p_1}},
\label{eqn:sf}
\end{equation}
where $C^x_p$ represents the percentage change of $x$ due the percentage change of a parameter $p$ from one stationary state to another.  

\begin{figure}[ht]
  \begin{center}
\subfigure[$K_M=1$]{\includegraphics[scale=0.6, angle=-90]{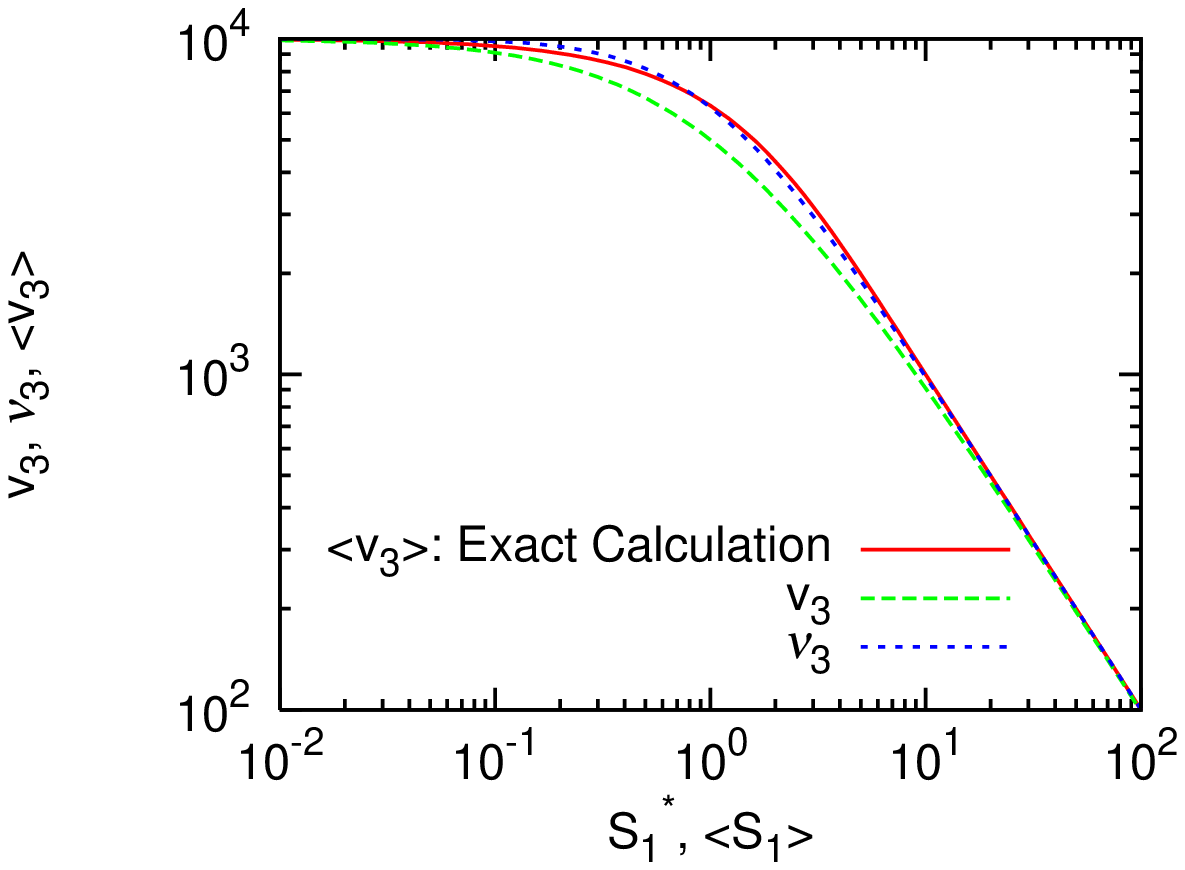}}
\subfigure[$K_M=0.1$]{\includegraphics[scale=0.6, angle=-90]{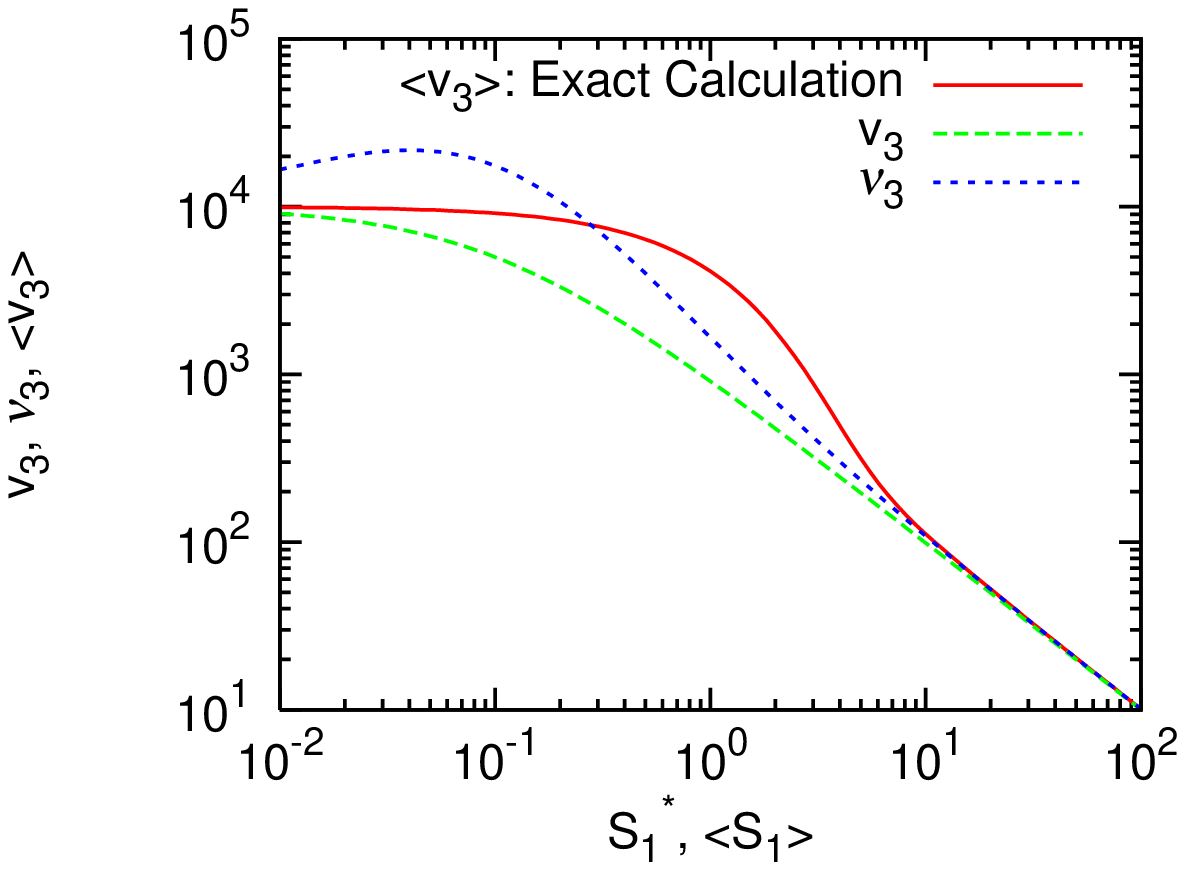}}
\subfigure[$K_M=0.01$]{\includegraphics[scale=0.6, angle=-90]{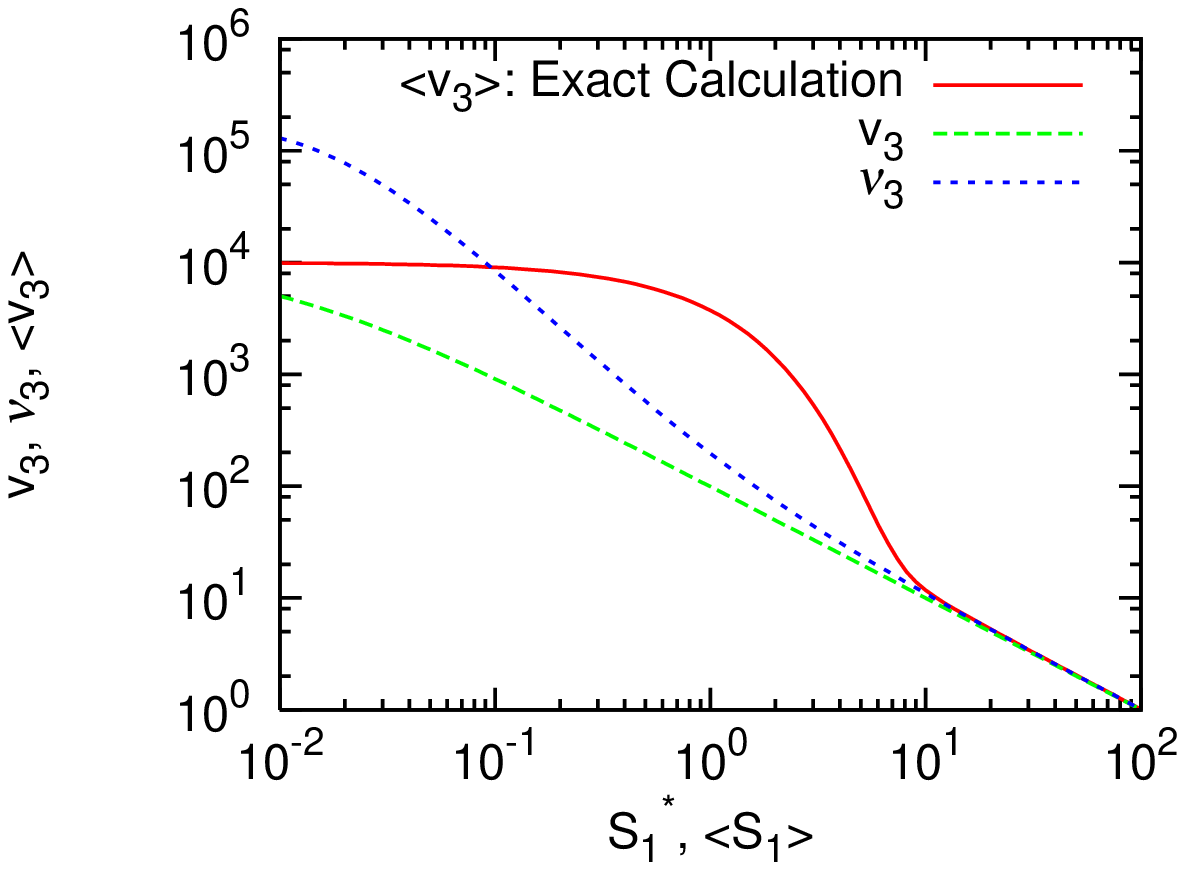}\label{fig:inhibition-c}}
  \end{center}
  \caption{ The stationary state degradation rate of $S_2$ vs. stationary state mean number of $S_1$ in a cascade reaction system as shown in Fig.~\ref{fig:sf}: $p_3=10000$, $K_M = 1, 0.1, 0.01$, and $p_1/p_2$ are varied from $1$ to $100$.}
  \label{fig:inhibition}
\end{figure}

\begin{figure}[ht]
\begin{center}
\subfigure[Sensitivity $C_{p_1}$ vs. mean number of $S_1$]{\includegraphics[scale=0.5, angle=-90]{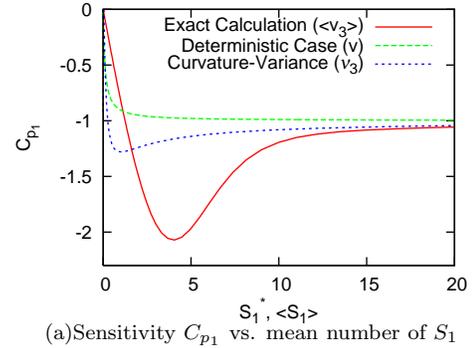}  \label{fig:SF}}\\
\subfigure[Propensity Functions]{\includegraphics[width=7cm, height=5cm]{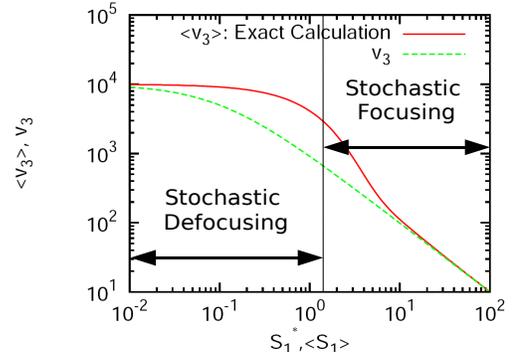} \label{fig:SF-arg} }
\caption{(a) Sensitivity control coefficiet (Eq.~\ref{eqn:sf}) vs. mean molecule number of $S_1$ for a cascade reaction system as shown in Fig.~\ref{fig:sf} for $K_M = 0.1$.  $p_1/p_2$ are varied from $1$ to $100$.  (b) Stochastic focusing-defocusing compensation: Mean values of a propensity function $v_3$ becomes larger than the estimate in the deterministic case  due to the curvature-variance effect.}
\end{center}
\end{figure}

The mean propensity $\nu_3$ becomes larger than the deterministic rate $v_3$ because the curvature of $v_3$ is positive.    Since $v_4$ is a linear function to $S_2$  (no curvature), $\nu_4(\langle s \rangle)$ is the same as $v_4 (\langle s \rangle)$.   In the stationary state, the mean degradation rate $\nu_4$ becomes balanced with $\nu_3$.   Thus, the mean level of $S_2$ will be enhanced as shown in Fig.~\ref{fig:inhibition}. The mean propensity function $\nu$ estimates the true mean rate of reaction quite accurately for the value of $K_M>1$.  However, for lower values of $K_M$ the correction becomes less accurate where the stochastic focusing can become significant.   This shows the limitation of our approximation.   We also show the above defined sensitivity coefficients for the case of $K_M=0.1$ in Fig.~\ref{fig:SF}.   The sensitivity is enhanced for the region of  $\langle  S_1 \rangle \gtrsim 1 $ while reduced for $\langle S \rangle \lesssim 1$.   Here stochastic focusing in one parameter region comes with stochastic de-focusing in another region.  This also can be understood by using the curvature-covariance correction effect in the mean propensity function  $\nu_3$.   At  $\langle S \rangle=0$, there is no variance and thus $\nu_3 = v_3$.  At $\langle S \rangle =\infty$, the hyperbolic curve becomes linear and its curvature vanishes, resulting in $\nu_3=v_3$.   As $\langle S \rangle$ decreases from $\infty$,  $\nu_3$  increases faster than $v_3$ but as $\langle S \rangle$ becomes closer to zero, $\nu_3$  increases slower than $v_3$ (see Fig.~\ref{fig:SF-arg}).  Thus, stochastic focusing and defocusing appears one after the other.    This implies that one can achieve higher sensitivities of one region by sacrificing sensitivities of another region.  We use this stochastic focusing-defocusing compensation effect to enhance the amplification of concentration detection designed by using incoherent feedforward networks, in the next example.   

The stochastic focusing we discuss has minor conceptual differences from the one discussed in Paulsson et al.~\cite{Paulsson2000}.  There stochastic focusing is the only effect of a rapidly fluctuating $S_1$.  However, here the stochastic focusing is independent of how fast $S_1$ fluctuates.   This difference is due to the fact that the focus  in Paulsson et al \cite{Paulsson2000} was on what is the most probable state of $S_2$, and we focus on the mean value of the $S_2$.    To understand this difference, we need to understand the dynamics of $S_2$ that is correlated with the fluctuation in $S_1$.  When all reaction rates $v_1$, $v_2$, $v_3$, and $v_4$ are in the same order of magnitude, $S_1$ and $S_2$ fluctuate on the similar time scales.  If $S_1$ hits zero the inhibition of $S_2$ is removed and thus the number of $S_2$ can rapidly increase to a very large number.  When $S_1$ increases to $1$, however,  the inhibition acting on $S_2$ appears and  the number of $S_2$ rapidly decreases.  Therefore, the time series profile of $S_2$ shows a flat lower bound at zero with many large sharp spikes.  This time series profile changes as the creation and degradation of $S_1$ get faster.  If the parameter values of $p_1$ and $p_2$ increase such that  $S_1$ fluctuates much faster than $S_2$,  $S_2$ sees the averaged behavior of $S_1$ and is unlikely  to hit zero.  Thus, the strong/weak inhibition by $S_1$ is averaged out.   Due to this averaging, the sensitivity enhancement, i.e., stochastic focusing, manifests itself.  This is why the stochastic focusing was claimed to be the effect of a rapidly fluctuating $S_1$ in this example \cite{Paulsson2000}.  However if one can observe the time series profile for a very long time to get good statistics of the spike heights,  the mean value of $S_2$ (the mean propensity $v_4$) becomes actually independent from  how fast $v_1$ and $v_2$ are, if the ratio $p_2/p_1$ is presumed to be kept constant.  This is why the stochastic focusing defined here becomes time-scale independent.

\subsection*{Example 4: Concentration Detection from Incoherent Feedforward Networks}
In this section we investigate how the stochastic focusing-defocusing compensation effect can be directly related to the sensitivity change of an incoherent feedforward network acting as a  concentration detector \cite{Entus2007}.   The stochastic focusing-defocusing compensation effect  will be shown to explain why stochastic fluctuation can amplify concentration detection while the sensitivity of detection is not enhanced.   

\begin{figure}[ht]
  \begin{center}
\subfigure[Poisson Distribution in $S_1$]{\includegraphics[scale=0.5]{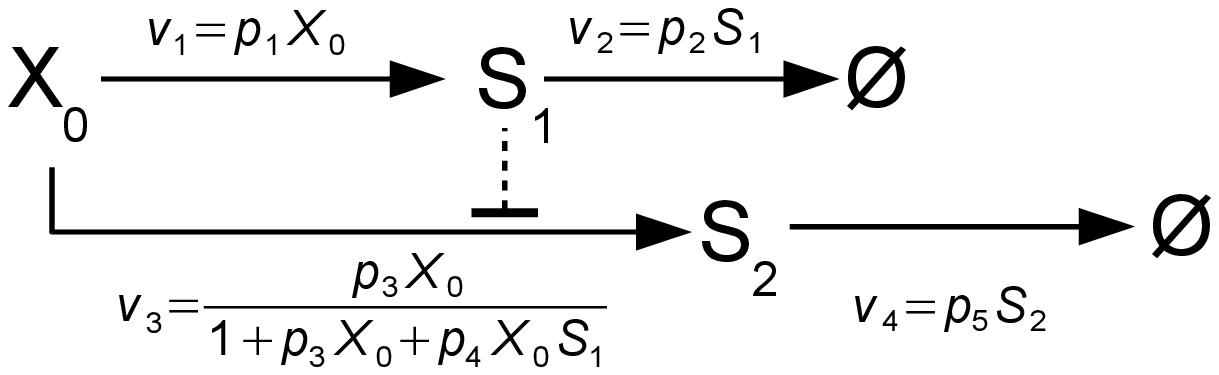}\label{fig:cd-a}}
\subfigure[Non-Poisson Distribution in $S_1$]{\includegraphics[scale=0.5]{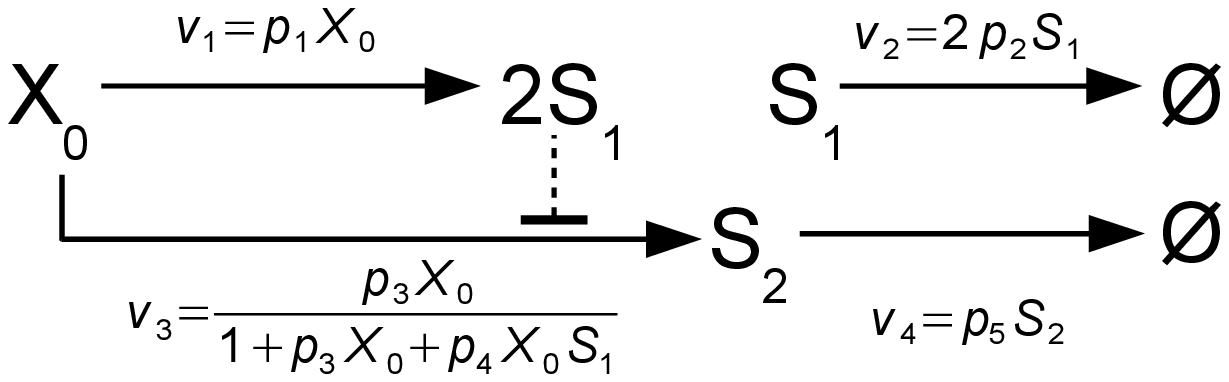}}
\subfigure[$X_0$ is allowed to fluctuate.]{\includegraphics[scale=0.5]{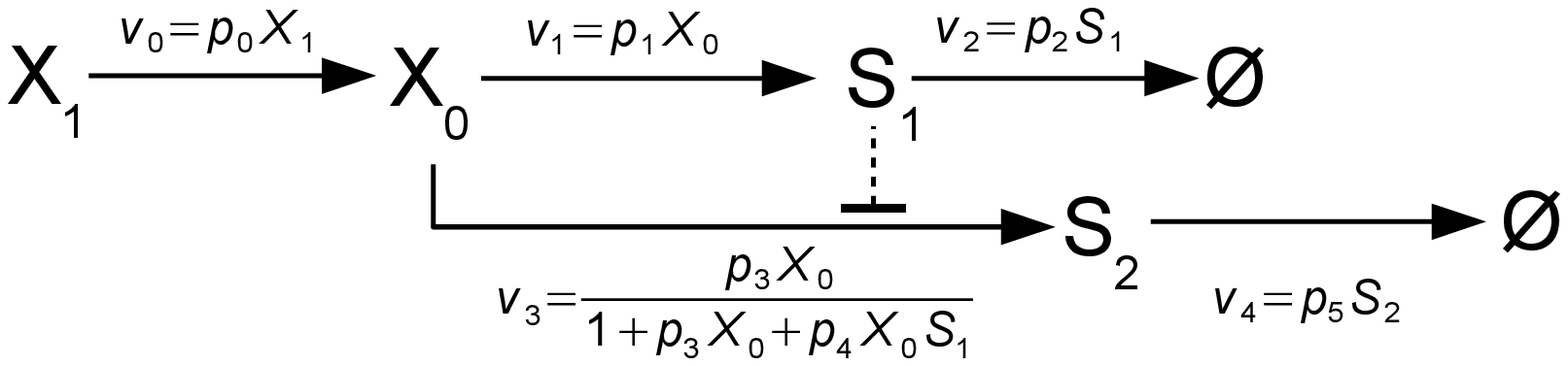}}
 \end{center}
  \caption{Incoherent feedforward reaction systems: $S_2$ is created from $X_0$ under the inhibition control of $S_1$ that also be created from $X_0$.  $X_0$ is a boundary species of which the concentration is not allowed to fluctuate in (a) and (b).  In (c), $X_0$ is allowed to fluctuate.  (a) The fluctuation in $S_1$ follows the Poisson distribution. (b) The fluctuation in $S_1$ is wider than the case (a) with the mean concentration of $S_1$ remaining the same.   (c) $X_0$ is allowed to fluctuate.   }
  \label{fig:ff-diag}
\end{figure}

\begin{figure}[h!]
\begin{center}
\subfigure[Concentration detection of $X_0$ by Observing $\langle S_2 \rangle$] {\includegraphics[scale=0.7, angle=-90]{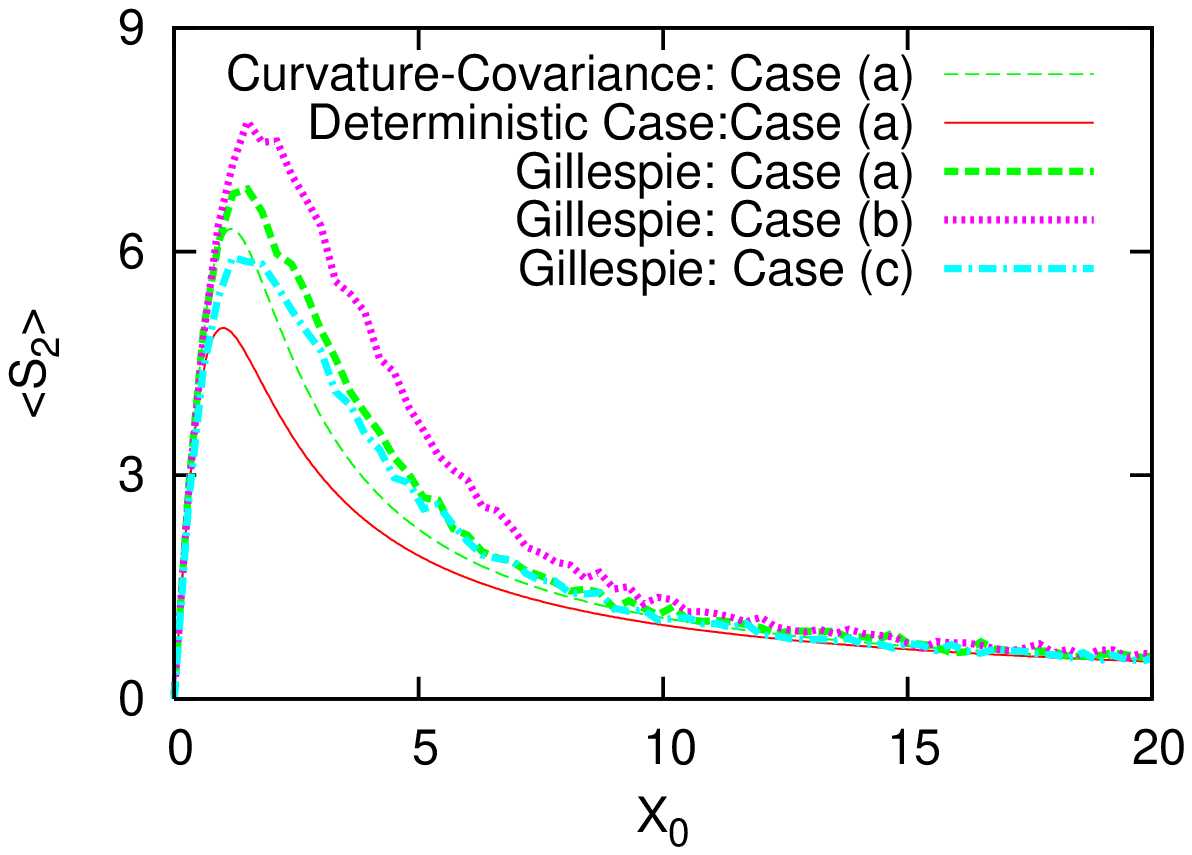}}
\subfigure[Significant Amplifications Due to Stochastic Effects]{\includegraphics[scale=0.7, angle=-90]{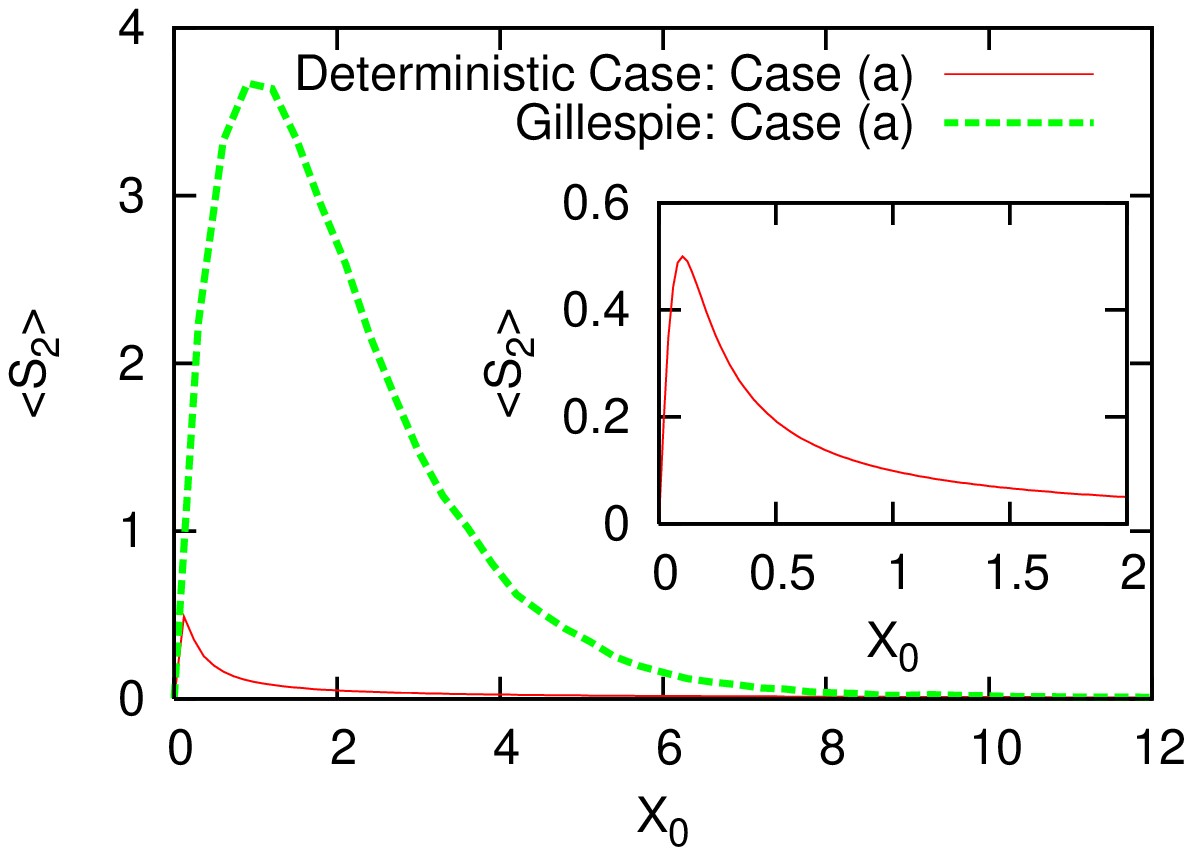}\label{fig:ff-huge}}
\caption{A Concentration Detector: mean concentration becomes very large only within a narrow region of the  value of $X_0$. Three different cases shown in Fig.~\ref{fig:ff-diag} are compared.  For the case (a), the estimates based on our analysis [Curvature-Covariance: Case(a)] is compared with its determinstic case and its stochastic simulations [Gillespie: Case (a)].     Parameters: $p_0=1$, $p_1=1$, $p_2=1$, $p_3=0.01$, $p_4=1$, $p_5=0.001$ for subfigure (a).  Significant amplification can be achieved more than 8 times of the deterministic case as shown in subfigure (b).  Parameters:   $p_1=1$, $p_2=1$, $p_3=0.01$, $p_4=100$, $p_5=0.001$ }
  \label{fig:ff}
\end{center}
\end{figure}

We explain first why incoherent feedforward networks can result in concentration detection.  As an example, consider a species ($S_2$) is regulated by two different pathways: either directly by $X_0$ or indirectly via $S_1$ as shown in Fig.~\ref{fig:ff-diag}.
The direct control acts as an activator for the production of $S_2$ while the indirect control acts as an inhibitor (see Fig.~\ref{fig:ff-diag}).  Thus, the feedforward is called incoherent.   When the concentration of $X_0$ is zero, $S_2$ is not created.  As $X_0$ increases, $S_2$ increases together but when $X_0$ becomes larger than a thresh-hold point it begins to decrease and eventually is dominated by $S_2$'s inhibition (Fig.~\ref{fig:ff}).  Thus, one can detect a specific range of the concentration of $X_0$ by monitoring the concentration of $S_2$.     

 As shown in Fig.~\ref{fig:ff} concentration detection is amplified  compared to the deterministic case, but sensitivities are not enhanced.   It is due to the stochastic focusing-defocusing compensation effect.  The networks represented in both Fig.~\ref{fig:sf} and  \ref{fig:cd-a} look identical  except that $S_1$ and $S_2$ are created from the common source in Fig.~\ref{fig:cd-a}.     Thus, stochastic focusing appears for large values of $\langle S_1 \rangle$, which means for large $X_0$, and stochastic defocusing for the smaller values.    In the case of Fig.~\ref{fig:cd-a}, the detection can be enhanced by  approximately 30\%  compared with  the deterministic case.   For a different parameter set as shown in Fig.~\ref{fig:ff-huge}, the detection can be enhanced by more than eight times.  This is because this parameter set gives a similar condition to the case of the cascade reaction network for $K_M=0.01$ as shown in Fig.~\ref{fig:inhibition-c}, where stochastic focusing  becomes significant.

To enhance the amplification, we exploit the curvature-covariance effect.  Since the curvature of $v_3$ with respect to $S_1$ is positive, stochastic focusing gets stronger  with the increase of the variance of $S_1$.    To increase  the variance, we replace the upstream reaction network of the creation and degradation reactions of $S_1$  as in Fig.~\ref{fig:ff-diag} (b): $X_0$ creates two $S_1$ molecules with the same reaction rate and degrades two times faster.  Thus, the mean values of the concentration of $S_1$ does not change but its fluctuation is shown to increase \cite{Elf2003}. This further amplifies the detection as Fig.~\ref{fig:ff}. 

We modify the original upstream network to increase the noise fluctuation of $S_1$ in a different way.  We allow $X_0$ to fluctuate, while both the mean concentrations of $X_0$ and $S_1$ are not changed (see Fig.~\ref{fig:ff-diag} (c)).   Although the variance of $S_1$ increases due to the noise propagated from $X_0$, the amplification is reduced.   This is because we have not taken into account the variance effect of $X_0$, which contributes  to the amplification negatively.  From Eq.~\ref{eqn:vtilde} the curvature-covariance correction term is given by
\[
\frac{1}{2}\frac{\partial^2 v}{\partial x_0^2} \sigma_{x_0,x_0}+ \frac{\partial^2 v}{\partial x_0 \partial s_1}\sigma_{x_0,s_1} + \frac{1}{2}\frac{\partial^2 v}{\partial s_1^2} \sigma_{s_1,s_1}
\]
The first (third) term is negative (positive) because the curvature is negative (positive) with respect to the change of $x_0$ ($s_1$).  The second term vanishes because the covariance between $X_0$ and $S_1$ vanishes.  The vanishing covariance happens to be true only in the stationary state \cite{Evans2005, Levine2007,KimdenNijs2007}.   Thus the amplification becomes smaller than the original case.

\section{Discussion and Conclusion}
\label{sec:conclusion}
We have analyzed the stationary state properties of stochastic reaction systems based on the mass fluctuation kinetics \cite{Gomez2007}, focusing on how intrinsic noise propagation affects system sensitivities.  We have considered nonlinear propensity functions such as Michaelis-Menten rate equations as a first step toward to understanding the relationship between network topologies and sensitivity changes.

We have investigated how mean levels of concentrations can be estimated by using the correction to the deterministic prediction by taking into account both  curvature effect of propensity function and concentration co-variances.   The curvature-covariance correction has been applied to predict stochastic focusing, concentration detection, and bistability (readers are referred to Appendix \ref{appendix:bistability})  qualitatively.   
Our analysis shows that stochastic focusing comes with stochastic de-focusing typically in the systems showing sigmoidal responses in the propensity functions and hyperbolic-type (zero order unltrasensitivity) negative feedback.  As an application of the compensation effect, we have investigated incoherent feedforward concentration detectors.  The exploitation of the stochastic focusing can leads to the amplification of the concentration detection.  The detection sensitivity is however not enhanced due to the stochastic de-focusing.   We further analyzed how the amplification is affected by additional intrinsic noise.   We have provided two cases that the amplification is enhanced and decreased  and have explained the behavior by using our curvature-covariance correction.    The upstream sub-network structure is shown to be significantly important to control the amplification.   We have also analyzed a bistable reaction system having a positive-feedback reaction network.  We have explained why intrinsic noise can reduce the bistability by using our analysis (see Appendix \ref{appendix:bistability}).  

As a further application of our analysis, we have shown that the stochastic version of MCA theorems exists:  these theorems can be derived by replacing $\bv$ in MCA theorems to $\bnu$.   We name this extension stochastic control analysis (SCA).  However, SCA has one drawback: elasticity (of a mean propensity function $\bnu$) does not become a local sensitivity measure any more.   This is because noise propagation can affect the mean propensity function.   In MCA, the global sensitivities (control coefficients) are expressed in terms of the local sensitivities (elasticities) and the system-wide response can be described by the combination of the local response.  However, in SCA such a description is not possible due to the non-local properties of the elasticities.   To resolve this issue, the noise propagation needs to be described by modular structures, i.e., local transfer functions \cite{Tanase2006, Warren2006, Simpson2003, Austin2006}.  We have been investigating the modular structure of the noise propagations as our current research to convert SCA into a useful form.  For a quantitative analysis, SCA needs to be applied to mass-action reaction systems without assuming quasi-steady state approximation \cite{Gomez2007}.
  
Our analysis is focused on the stationary state static responses.  However, it can be extended for the study of dynamic responses by taking Fourier transformations of Eqs.~\ref{mean1} and \ref{variance1} just like dynamic  response studied in classic metabolic control analysis in \cite{Ingalls2004, Rao2004}.    One of the applications of the dynamic response analysis is feedforward networks acting as frequency filters.

We hope that the curvature-covariance effect and the stochastic focusing-defocusing compensation effect can be used for intuitive understanding of how stochastic intrinsic noise affects the system functions.  This understanding will help to  systematically  design and improve a synthetic genetic circuit \cite{Entus2007, Bashor2008}, especially how to control the mean levels of concentrations by modifying subnetworks in stationary states.

\section*{Acknowledgement}
We are very grateful to Mustafa Khammash, Johan Paulsson, Sukjin Yoon, and our group members: Frank T. Bergmann, Deepak Chandran, Sean Sleight,  and Lucian Smith for useful discussions and comments.   We acknowledge the generous support from University of Washington and some support from the National Science Foundation in this work (NSF 0527023-FIBR).


\begin{appendix}

\section{The Chemical Master Equation}
Chemical reactions occur in a random fashion due to collisions between reactants.  We assume that each reaction event arises homogeneously (in a uniform fashion in position space) and also independently to another event.  Such reaction systems are often modeled by stochastic processes.   The processes are  fully described by the time evolution of probability distribution of reactant number for each species at a given time, mathematically formulated by the chemical master equation \cite{Gillespie1992}.   This equation describes the time evolution of a probability distribution function, which represents the probability that one finds the number of molecules for each species at a given time. 


We consider $m$ species of molecules involved in $n$ reactions: 
\begin{eqnarray*}
n_1^l S_1 + \cdots + n_m^l S_m \xrightarrow{V_l} m_1^l S_1 + \cdots + m_m^l S_m,
\end{eqnarray*}
where the molecule numbers are denoted by $\{S_i\}$ with $i=1, \cdots, m$ and the rate of reaction by $V_l$.  We also assume that $V_l$ can be controlled by changing a parameter $p_l$.  The number change in species $i$ by a single event of the above reaction $l$ is described by a reduced stoichiometry matrix: $N_{R_{il}} \equiv m_i^l-n_i^l$.    The numbers evolve stochastically and their evolutions can be described by the chemical master equation:  
\begin{eqnarray}
\frac{\partial P(\bS,t)}{\partial t}& =& \sum_{j=1}^n \Big[ P(\{S_i-N_{R_{ij}}\},t)V_j(\{S_i-N_{R_{ij}}, p_j\}) \nonumber\\
&& - P(\bS,t)V_j(\bS,p_j) \Big].
\label{app:master}
\end{eqnarray}
The first term in the right hand side corresponds to the probability increase due to the state change: $\{ S_i -N_{R_{ij}} \} \rightarrow \bS$ through events of reaction $j$ occurred at time $t$.  The second to the probability decrease due to the state change: $ \bS\rightarrow \{ S_i +N_{R_{ij}} \}$.

\section{Derivations of Eqs.~\ref{mean1} and \ref{variance1}}
\label{appendix-mean-variance}
We switch the representation of states from numbers $\{ \bS \}$ to concentrations of  molecules $\{\bs \equiv \bS/\Omega \}$ with $\Omega$ a system volume, since this concentration representation has a direct correspondence to deterministic macroscopic kinetics.  

The mean concentration of a species $i$  is given as $\langle s_i \rangle \equiv \big[\prod_{j=1}^m \int_{s_j=0}^ \infty ds_j\big] s_i P(\bs,t)$ where $P(\bs,t)$ is the probability distribution function of molecule concentrations, $\bs= \{s_1, s_2, \cdots, s_m\}$.  The evolution of the mean concentration of a species $i$ is shown later in this section to be governed by the following equation:
\begin{equation}
\frac{d\langle \bs \rangle }{dt} = \bN_{R}  \langle \bv(\bs,\bp) \rangle, 
\label{app:mean}
\end{equation}
 where $\bN_R$ is a reduced stoichiometry matrix and $\bv$ represents propensity functions: $\bv\equiv \{ v_1, \cdots, v_n\}$ with $v_i \equiv V_i/\Omega$.   We note that $v_i$ is a function of $\{s_j\}$ with $j=0, \cdots, m_0$ and $p_i$, where  $m_0$ is the number of linearly independent rows in a stoichiometry matrix.

A concentration covariance between two species ($i$ and $j$) is defined as 
\[
\sigma_{ij} \equiv \Big\langle (s_i -\langle s_i \rangle )(s_j -\langle s_j \rangle )\Big\rangle.
\]
The correlations between different molecular species $i$ and $j$ and the variance of a species $i$ are quantified by $\sigma_{ij}$ and $\sigma_{ii}$, respectively.  The correlation measures the statistical independence between two random fluctuations in each different species.  If the correlation vanishes, their concentration fluctuations are independent statistically.  E.g., if $\sigma_{ij} = 0$ with $i \neq j$, the fluctuation with respect to the mean value of $s_i$ statistically is not related to the fluctuation of $s_j$, although the mean value of $s_i$ may change depending on  the mean value of $s_j$ through Eq.~\ref{app:mean}.  The variance of $s_i$, $\sigma_{ii}$, quantifies its fluctuation strength.  The evolution of the covariance matrix is shown later in this section to be described by the following equation: 
\begin{equation}
\frac{d \bsigma  }{dt} = \Big\langle (\bN_R \bv)  (\bs-\langle \bs \rangle) + (\bs-\langle \bs \rangle)^T  (\bN_R \bv)^T + \frac{\bD}{\Omega} \Big \rangle,
\label{app:variance}
\end{equation}
where the diffusion coefficient matrix $\bD$ is defined by $ \bN_{R} \bLambda \bN_{R}^T $ with a diagonal matrix $\Lambda_{ij} \equiv v_i\delta_{ij}$. 

It  is almost impossible to solve equations~(\ref{app:mean}) and (\ref{app:variance}) unless the propensity function $\bv(\bs,\bp)$ is linear with $\bs$.  E.g., for a nonlinear function $v(s,e)=s^2$, equation~\ref{app:variance} cannot be solved unless the third moment of $s-\langle s \rangle$, $\langle (s-\langle s \rangle )^3 \rangle$, is known already.  To find the value of  the third moment, the fourth moment needs to be known too.  The same argument applies to all the higher moments.  Thus, we need to truncate the moment series to solve these equations \cite{Kampen2001}.  In the following paragraphs,  we assume that the third and higher order moments are neglected, and then we derive Eqs.~\ref{mean1} and \ref{variance1}, which describe the approximate evolutions of the mean values and covariances of the concentrations \cite{Gomez2007}.   

First,  equations~\ref{app:mean} and \ref{app:variance} are derived from the master equation, Eq.~\ref{app:master}.   Equation~\ref{app:mean} is derived as follows.
\begin{eqnarray*}
\lefteqn{\frac{d \langle S_k  \rangle }{d t}= \sum_S S_k \sum_{j=1}^R \Big[  \Big( \prod_{i=1}^m E^{-N_{R_{ij}}}\Big) -1    \Big]V_j P}\\
	&=& \sum_S  \sum_{j=1}^n \Big[ (S_k- N_{R_{kj}})  \Big( \prod_{i=1}^m E^{-N_{R_{ij}}}\Big) V_j P - S_kV_j P \Big]\\
	&&+ \sum_S \sum_{j=1}^n N_{R_{kj}}\Big( \prod_{i=1}^m E^{-N_{R_{ij}}}\Big) V_j P\\
	&=& \sum_S  \sum_{j=1}^n \Big[ \Big( \prod_{i=1}^m E^{-N_{R_{ij}}}\Big) S_k V_j P - S_kV_j P \Big]\\
	&&+ \sum_S \sum_{j=1}^n N_{R_{kj}} V_j P, 
\end{eqnarray*}
where $E^{-N_{R_{ij}}}$ is a raising/lowering operator: $E^{-N_{R_{ij}}}f(\bS) = f(\{  S_{ij} - N_{R_{ij}}  \})$. 
The first term in the right hand side vanishes.  We now switch this number representation to the concentration representation by replacing $\bS \rightarrow \Omega \bs$ and $\bV \rightarrow \Omega \bv$.  Then,  equation~\ref{app:mean} is derived after the cancellation of $\Omega$ in both hand sides of the above equation.

Equation~\ref{app:variance} is derived as follows.  First we define, a number covariance matrix $\bSigma$:
\[
\Sigma_{ij} \equiv \Big\langle (S_i - \langle S_i \rangle) ( S_j - \langle S_j \rangle) \Big\rangle.
\]
The time evolution of this covariance matrix is given by:
\begin{widetext}
\begin{eqnarray*}
\frac{d \langle \Sigma_{kl} \rangle}{dt} &=& \sum_{\bS} (S_k -\langle S_k \rangle )(S_l -\langle S_l \rangle )  \sum_{j=1}^n \Big[  \Big( \prod_{i=1}^m E^{-N_{R_{ij}}}\Big) -1    \Big]V_j P\\
	&=& \sum_S  \sum_{j=1}^n \Big[  \Big( \prod_{i=1}^m E^{-N_{R_{ij}}}\Big)(S_k -\langle S_k \rangle +N_{R_{kj}})(S_l -\langle S_l \rangle +N_{R_{lj}})-(S_k -\langle S_k \rangle )(S_l -\langle S_l \rangle )    \Big]V_j P\\
	&=& \sum_{j=1}^n\Big \langle (S_k -\langle S_k \rangle +N_{R_{kj}})(S_l -\langle S_l \rangle +N_{R_{lj}}) V_j-  (S_k -\langle S_k \rangle )(S_l -\langle S_l \rangle )V_j\Big \rangle \\
	&=& \sum_{j=1}^n \Big\langle (S_k -\langle S_k \rangle)N_{R_{lj}}  V_j + V_j N_{R_{kj}} (S_l -\langle S_l \rangle ) + N_{R_{kj}}N_{R_{lj}} V_j \Big \rangle \\
	&=& \Bigg\langle \Big[  (\bS - \langle \bS \rangle)^T (\bN_R \bV)^T +  (\bN_R \bV)(\bS - \langle \bS \rangle)  + \bN_R \bLambda^\prime \bN_R^T  \Big] \Bigg \rangle \Bigg|_{kl},
\end{eqnarray*}
\end{widetext}
where $\Lambda^\prime_{ij} \equiv V_i \delta_{ij}$.   
By switching the number representation to the concentration representation ($\bSigma = \Omega^2 \bsigma$), we derive Eq.~\ref{app:variance}.

Equations~\ref{mean1} and \ref{variance1} are derived from Eqs.~\ref{app:mean} and \ref{app:variance} by using the Tayler expansion of  the propensity function  $\bv$ with respect to $\langle \bs \rangle$ and by neglecting the third and higher moments.  We will discuss this approximation further in Appendix~\ref{appendix-truncation}.  Then,  equation~\ref{app:mean} can be expressed as
\begin{eqnarray*}
\lefteqn{ \frac{d\langle \bs \rangle }{dt}  =  \bN_{R} { \Bigg \langle}   \bv(\langle \bs \rangle ,\bp) \rangle  + \sum_{i=1}^{m_0} \frac{\partial \bv}{\partial s_i} \Big |_{\bs = \langle \bs \rangle} (s_i - \langle s_i \rangle )}\\
	&&+ \frac{1}{2} \sum_{i,j=1}^m \frac{\partial^2 \bv}{\partial s_i \partial s_j} \Big|_{\bs = \langle \bs \rangle} (s_i - \langle s_i \rangle) (s_j - \langle s_j \rangle)  \Bigg \rangle.
\end{eqnarray*}
The second term in the right hand side vanishes because $\langle s_i - \langle s_i \rangle \rangle = \langle s_i \rangle - \langle s_i \rangle =0$.  Equation~\ref{mean1} is derived.  

In the similar way, Eq.~\ref{variance1} also can be derived from Eq.~\ref{app:variance}.  The propensity function $\bv$ is expanded by using the Taylor expansion.  Then, the first term in the right hand side of Eq.~\ref{app:variance} becomes, after neglecting the third and higher moments:
\[
\big \langle (\bN_R \bv)(\bs - \langle \bs \rangle) \big\rangle = (\bN_R \bv(\langle \bs \rangle )\big\langle(\bs - \langle \bs \rangle) \big\rangle + \bJ \bsigma,
\]
where $\bJ$ is a Jacobian matrix defined as $J_{ij} \equiv \sum_k N_{R_{ik}}\partial v_k / \partial s_j |_{\bs = \langle \bs \rangle}$.  The first term in the above vanishes.     We take the same procedure for the second term in the right hand side of Eq.~\ref{app:variance}.   The third term of Eq~\ref{app:variance} has an extra factor of $1/\Omega$ and we neglect the terms of the order of $1/\Omega$  and the higher in the Taylor expansion of the propensity function which comprises the diagonal elements of $\bLambda$:
\[
\langle \Lambda_{ij} \rangle  = \delta_{ij} \langle v_i \rangle = \delta_{ij} \Big[ v_i(\langle \bs \rangle) + \frac{\partial v_i}{\partial \bs}\Big|_{\bs = \langle \bs \rangle}\big\langle (\bs-\langle \bs \rangle) \big\rangle   \Big].
\]
The second term vanishes.  Therefore, equation~\ref{variance1} is derived.

\section{Moment Closure Approximations}
\label{appendix-truncation}
In this section, we will explain in detail the approximation we have taken to derive Eqs.~\ref{mean1} and \ref{variance1}.      The first assumption that we have taken is that intrinsic noise is  not strong enough that the propensity function $v$ can be expanded in Taylor series.   The Taylor expansion of $v(s)$ with respect to $\langle s \rangle$ should converge in the region that the value of $s$ is sampled with high probability.  The second assumption is that the third and higher moments of $s-\langle s \rangle$ are negligible.   We will explain the second assumption in detail in the following  paragraphs. 

We consider a simple reaction system: 
\[
 \xrightarrow{k_0} S \xrightarrow{v(s)},
\]
where the creation rate is a constant.   The mean propensity of $v$ is Taylor-expanded and the second and third terms of the Taylor expansion are compared with each other.  The third is required to be much smaller than the second for the approximation to be valid:
\[
\frac{1}{2}\frac{\partial^2 v}{\partial s^2}\Big|_{s=\langle s \rangle} \sigma\gg \frac{1}{3!}\frac{\partial^3 v}{\partial s^3}\Big|_{s=\langle s \rangle} \langle (s-\langle s \rangle)^3 \rangle.
\]
If the mean concentration of $s$ is fixed while a system volume $\Omega$ increases,  $v(\langle s \rangle)$ is invariant under the change of the system volume.  Thus, the inequality in the above is determined by $\langle (s-\langle s \rangle)^2 \rangle$ and $\langle (s-\langle s\rangle)^3 \rangle$.   Then, why can it be reasonable that $\langle (s-\langle s\rangle)^3 \rangle$ is much smaller than $\langle (s-\langle s \rangle)^2 \rangle$ for the large system volume?  We assume that a system is near the stationary state.  Otherwise, the transient process to the stationary state can be arbitrary because the process depends on the initial condition of the process.   If $v$ is almost a linear function of $s$, the distribution of $s$ becomes similar to the Poisson distribution.  Then, the higher the moments the smaller in magnitude; 
\begin{eqnarray*}
\langle s \rangle &=& \langle S \rangle / \Omega ={\cal O}(1),\\ 
\langle (s-\langle s \rangle )^2 \rangle  &=& \langle (S-\langle S \rangle)^2  \rangle / \Omega^2 \simeq \langle S \rangle / \Omega^2={\cal O}(1/\Omega), \\
\langle (s-\langle s \rangle )^3 \rangle  &=& \langle (S-\langle S \rangle)^3  \rangle / \Omega^3 \simeq \langle S \rangle / \Omega^3={\cal O}(1/\Omega^2),\\
\langle (s-\langle s \rangle)^4 \rangle &\simeq&(  \langle S \rangle  + 3\langle S \rangle^2)/\Omega^4={\cal O}(1/\Omega^2).
\end{eqnarray*}
If $\Omega$ is large enough that the above criteria satisfied, then the moment closure approximation is justified for the process similar to the Poisson process.  If $v$ is highly nonlinear in $s$, the above scaling relationship does not hold any longer.  If we assume the fluctuation of $s$ is small enough that $v$ can be linearized with respect to the mean concentration of $s$,  the distribution of $s$ becomes similar to the Gaussian distribution.  For the Gaussian distribution, one can show the relationship between the second and fourth moments: $\langle (s-\langle s \rangle )^4 \rangle = 3 \langle (s-\langle s \rangle )^2 \rangle^2$.  This means that the fourth moment becomes much smaller than the second moment if the second moment is very small, and the moment closure approximation can become valid. 

There is another way of power counting.
Consider the case that the system volume $\Omega$ is fixed while the molecule number increases and the stochastic variable $S$ follows the Poisson process.  If the degree of the moment of $(S-\langle S \rangle)/\langle S \rangle$ gets higher, the magnitude of the moment becomes lower; 
\begin{eqnarray*}
\langle (S-\langle S \rangle )^2 \rangle/\langle S \rangle^2  &=& 1/\langle S \rangle, \\ 
\langle (S-\langle S \rangle )^3 \rangle/\langle S \rangle^3  &=& 1/\langle S \rangle^2,\\
\langle (S-\langle S \rangle )^4 \rangle/\langle S \rangle^4  &=&1/\langle S \rangle^3  + 3/\langle S \rangle^2.
\end{eqnarray*}  $(S-\langle S \rangle )/\langle S \rangle$ means the percentage change in the molecule number from the mean value. Thus it is more directly applicable to biological experiments and also simulations based on the Gillespie algorithm.  For example, consider in GFP experiments one wants to study the protein number statistics.  Then, the percentage change in GFP's can be estimated by measuring the light intensity fluctuation.  In Monte-Carlo simulations using the Gillespie algorithm, the simulation does not have volume as its parameter;  the volume information is rather hidden in the rate of reaction or in the reaction time scale.  

For the mathematical ease, we have used the former way of power counting to derive the Eq.~\ref{mean1} and \ref{variance1} (see Appendix \ref{appendix-mean-variance}).  To compare with Monte-Carlo simulations using the Gillespie algorithm,  we switch back to the number representations to compute the means and covariances in molecule numbers.  Thus, the latter way of power counting applies for Monte-Carlo simulation results.

\section{Bistability from a Positive Feedback Network}
\label{appendix:bistability}
In this section we investigate how a bistable system is affected  by concentration fluctuations.   We show that the concentration fluctuations can  reduce the bistability due to the curvature-covariance effect.  

The mass fluctuation kinetics (MFK) has been applied to a bistable system in G\'{o}mez-Uribe et al.~\cite{Gomez2007}.  MFK describes the time evolution of the system variables (mean and covariance values of concentrations) deterministically and it cannot describe stochastic switching: jumping from one local stable state to another.   However, the existence of the bistability can be determined by solving Eq.~\ref{mean2}, although in a very qualitative level.   When concentrations fluctuate around only one of the locally stable regions for most of the time and rarely jumps to the other stable one, Eq.~\ref{mean2} can provide a correction due to the local concentration fluctuation (however, without taking into account the fluctuations between the two local stable regions).   

\begin{figure}[h!]
\begin{center}
\subfigure{\includegraphics[scale=0.4, angle =0]{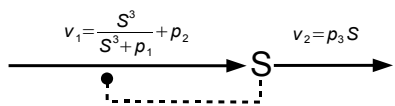}}
\caption{A positive feedback reaction system: $S$ itself accelerates the creation of itself and this is represented by  the Hill function.}
  \label{fig:pfb}
\end{center}
\end{figure}

\begin{figure}[h!]
\begin{center}
\subfigure[Probability Distribution of $S$]{\hspace{-0.2in}\includegraphics[scale=0.53, angle=-90]{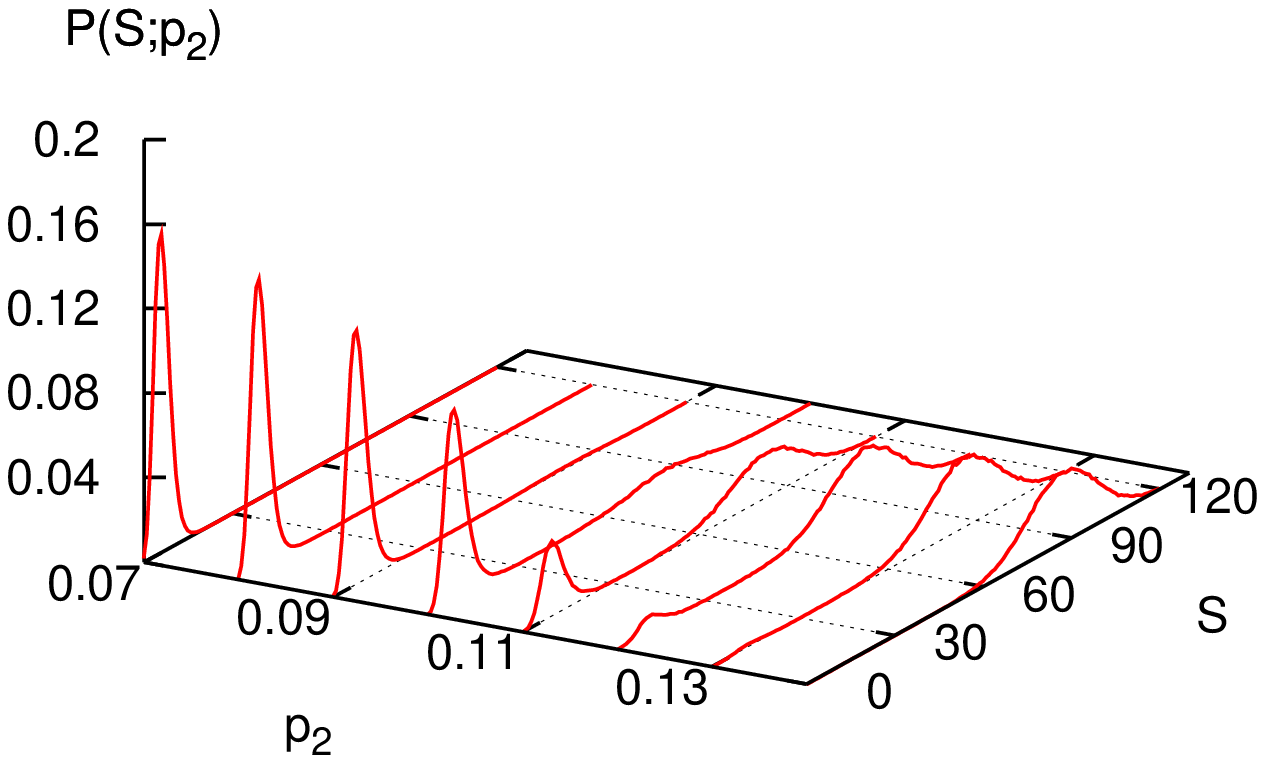}}
\subfigure[Mean Concentration]{\includegraphics[scale=0.53, angle=-90]{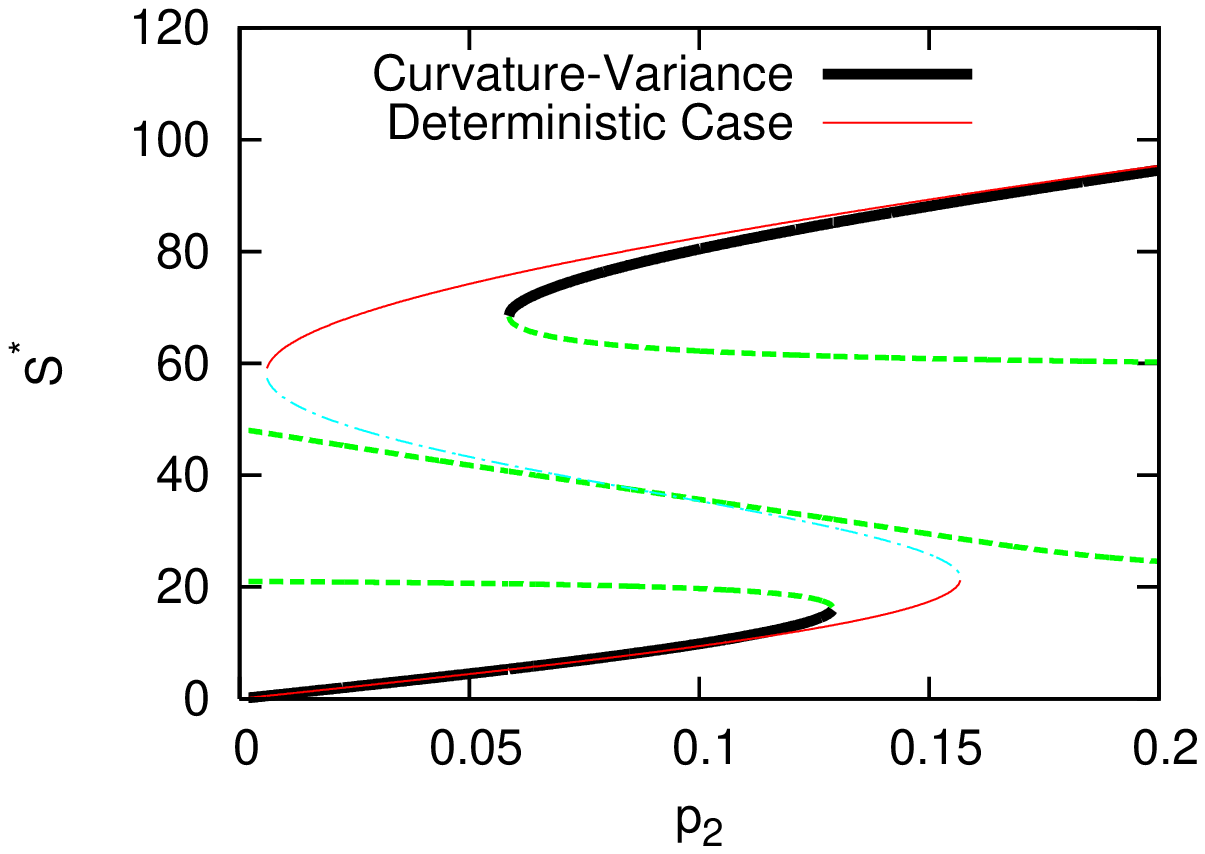}}
\subfigure[Propensity Functions]{\includegraphics[scale=0.53, angle=-90]{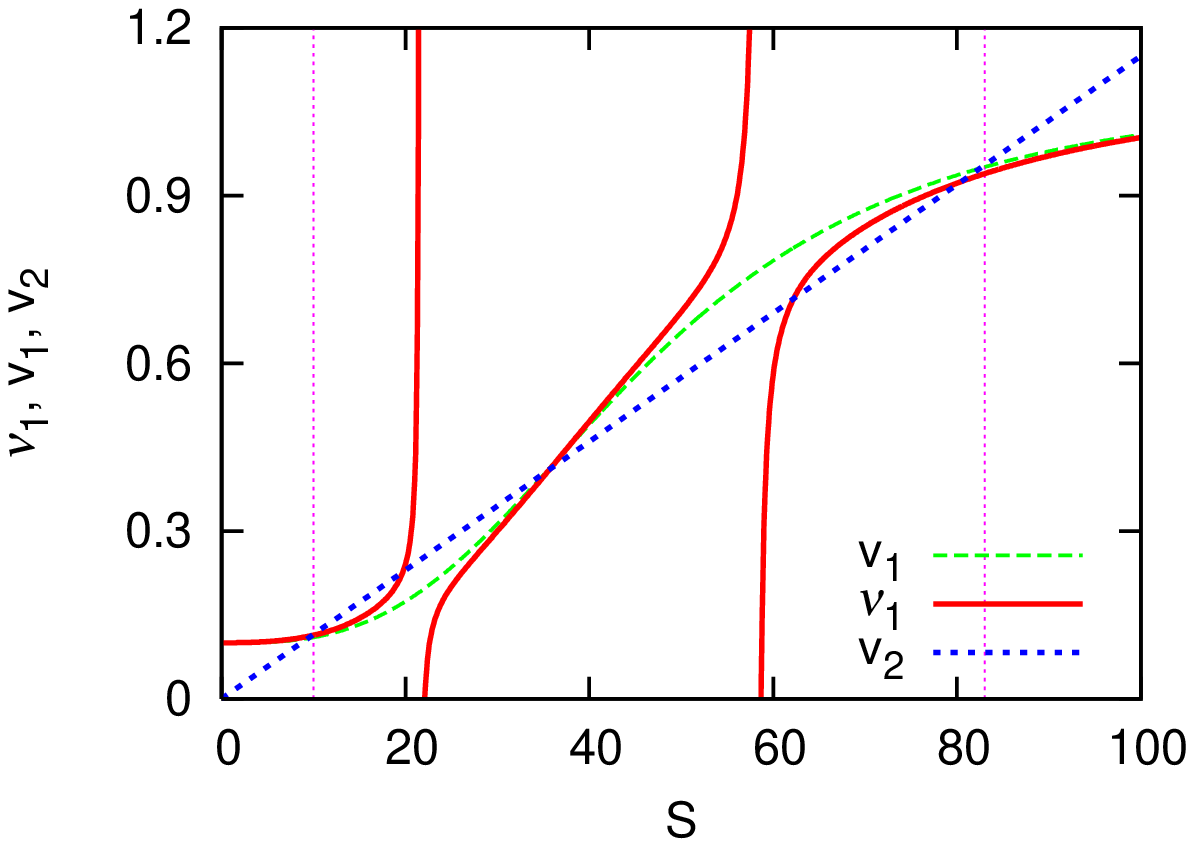}\label{fig:pfb-v}}
\caption{Bistability of a positive feedback system shown in Fig.~\ref{fig:pfb}:   The probability distribution function for $S$ shows double peaks depending on the value of $p_2$ in (a).   Stationary state values of  $S^*$ are estimated from each individual  deterministic and stochastic approaches in (b), where solid lines correspond to stable stationary stationary state concentrations and dotted lines to unstable ones.    In (c), the propensity function, $\nu_1$, diverges. Two purple lines indicate the  stable concentrations.  (Parameters: $p_1, p_3 = 100000,  0.0115$)  }
  \label{fig:pfb-s}
\end{center}
\end{figure}

We consider a positive feedback reaction network as shown in Fig.~\ref{fig:pfb}.   In the deterministic case,  the stationary state value of $S$ can take two different stable levels.  One corresponds to a  low number state and the other to a  high number state.   Now, we take into account the concentration fluctuation and  we examine how the fluctuation affects  bistability.   In the stochastic framework, bistability means the existence of double peaks in the concentration probability distribution function (see Fig.~\ref{fig:pfb-s}(a)).    There is always a finite probability  that the number $S$ meanders  back and forth between the low and the high number peaked regions (stochastic switching).   If the two peaks are well separated, such switching however rarely happens.  In this case, the  probability distribution function of $S$ can be considered a superposition of two of each mono-stable distribution functions centered at each peak.    Each distribution gives an estimate of the mean concentration level of one of the bistable states.   This is why our analysis based on the curvature-covariance effect can estimate when the system becomes bistable, although in a qualitative level  since the stochastic switching is neglected.    In this example as shown in Fig.~\ref{fig:pfb},  the curvature-variance correction reduces the bistability; $\nu_1$ larger than $v_1$ for the positive curvature region of $v_1$ and also makes $\nu_1$ smaller than $v_1$ for the negative curvature region.  Thus, bistability appears at the larger value of $p_2$ (see Fig.~\ref{fig:pfb}) and disappears at its smaller value as shown in Fig.~\ref{fig:pfb-s}(b).   However, there exist singular points in $\nu_1$ because $\sigma^*$ (the solution of Eq.~\ref{variance2}) is undetermined when the local slope of $v_1$ and $v_2$ are the same (see Fig.~\ref{fig:pfb-v}).   For our analysis to be valid, the mean concentration estimates (the solution of Eq.~\ref{mean2}) needs to be far away from the singular points.  Such singular behavior explains why more than one  unstable stationary level of $S$ is predicted as shown in Fig.~\ref{fig:pfb-s}(b).  The prediction of the unstable stationary levels should be neglected because the approximation used in deriving Eq.~\ref{variance2} becomes invalid. 


We have shown that intrinsic noise can reduce bistability resulting from a positive feedback by considering the curvature-covariance effect.   Although this analysis based on the curvature-covariance effect needs to be performed with a significant care when applied to a bistable system, a qualitative and intuitive level of understanding will be helpful to predict the stochastic noise effect on the system.





\end{appendix}

\bibliography{sensitivity}

\begin{thebibliography}{10}

\bibitem{footnote1}
The mathematical task of analyzing the chemical master equation is essentially
  the same as that of the Schr{\"{o}}dinger's equation in quantum chemistry.
  See H. Qian and D.A. Beard, {\em Chemical Biophysics: Quantitative Analysis
  of Cellular Systems.} Cambridge Texts in Biomedical Engineering, Cambridge
  University Press (2008).

\bibitem{Austin2006}
D.~W. Austin, M.~S. Allen, J.~M. McCollum, R.~D. Dar, J.~R. Wilgus, G.~S.
  Sayler, N.~F. Samatova, C.~D. Cox, and M.~L. Simpson.
\newblock Gene network shaping of inherent noise spectra.
\newblock {\em Nature}, 439:608--611, Feb 2006.

\bibitem{Bashor2008}
C.~J. Bashor, N.~C. Helman, S.~Yan, and W.~A. Lim.
\newblock Using engineered scaffold interactions to reshape map kinase pathway
  signaling dynamics.
\newblock {\em Science}, 319(5869):1539--1543, Mar 2008.

\bibitem{Elf2003}
J.~Elf and M.~Ehrenberg.
\newblock Fast evaluation of fluctuations in biochemical networks with the
  linear noise approximation.
\newblock {\em Genome Res.}, 13(11):2475--2484, Nov 2003.

\bibitem{Elowitz2002}
M.~B. Elowitz, A.~J. Levine, E.~D. Siggia, and P.~S. Swain.
\newblock Stochastic gene expression in a single cell.
\newblock {\em Science}, 297(5584):1183--1186, Aug 2002.

\bibitem{Entus2007}
R.~Entus, B.~Aufderheide, and H.~M. Sauro.
\newblock Design and implementation of three incoherent feed-forward motif
  based biological concentration sensors.
\newblock {\em Syst. Synth. Biol.}, 1:119--128, 2007.

\bibitem{Evans2005}
M.~R. Evans and T.~Hanney.
\newblock Nonequilibrium statistical mechanics of the zero-range process and
  related models.
\newblock {\em J. Phys. A: Math. Gen.}, 38:R195--R240, 2005.

\bibitem{Fell1992}
D.~A. Fell.
\newblock Metabolic control analysis: a survey of its theoretical and
  experimental development.
\newblock {\em Biochem. J.}, 286:313--330, Sep 1992.

\bibitem{Fell1996}
D.~A. Fell.
\newblock {\em Understanding the Control of Metabolism.}
\newblock London, Portland Press, 1996.

\bibitem{Gillespie1977}
D.~T. Gillespie.
\newblock Exact stochastic simulation of coupled chemical reactions.
\newblock {\em J. Phys. Chem.}, 81:2340--2361, 1977.

\bibitem{Gillespie1992}
D.~T. Gillespie.
\newblock A rigorous derivation of the chemical master equation.
\newblock {\em Physica A}, 188:404--425, 1992.

\bibitem{Gomez2007}
C.~A. G\'{o}mez-Uribe and G.~C. Verghese.
\newblock Mass fluctuation kinetics: capturing stochastic effects in systems of
  chemical reactions through coupled mean-variance computations.
\newblock {\em J. Chem. Phys.}, 126(2):024109, Jan 2007.

\bibitem{Goutsias2005}
J.~Goutsias.
\newblock Quasiequilibrium approximation of fast reaction kinetics in
  stochastic biochemical systems.
\newblock {\em J. Chem. Phys.}, 122(18):184102, May 2005.

\bibitem{Haseltine2002}
E.~L. Haseltine and J.~B. Rawlings.
\newblock Approximate simulation of coupled fast and slow reactions for
  stochastic chemical kinetics.
\newblock {\em J. Chem. Phys.}, 117:6959, 2002.

\bibitem{Hooshangi2005}
S.~Hooshangi, S.~Thiberge, and R.~Weiss.
\newblock Ultrasensitivity and noise propagation in a synthetic transcriptional
  cascade.
\newblock {\em Proc. Natl. Acad. Sci. U.S.A.}, 102(10):3581--3586, Mar 2005.

\bibitem{Ingalls2004}
B.~P. Ingalls.
\newblock A frequency domain approach to sensitivity analysis of biochemical
  networks.
\newblock {\em J. Phys. Chem. B}, 108:1143--1152, 2004.

\bibitem{Ingalls2006}
B.~P. Ingalls.
\newblock Metabolic control analysis from a control theoretic perspective
  decision and control.
\newblock In {\em 45th IEEE Conference}, pages 2116--2121, 2006.

\bibitem{Ingalls2003}
B.~P. Ingalls and H.~M. Sauro.
\newblock Sensitivity analysis of stoichiometric networks: an extension of
  metabolic control analysis to non-steady state trajectories.
\newblock {\em J. Theor. Biol.}, 222(1):23--36, May 2003.

\bibitem{Kacser1973}
H.~Kacser and J.~A. Burns.
\newblock The control of flux.
\newblock {\em Symp. Soc. Exp. Biol.}, 27:65--104, 1973.

\bibitem{KimdenNijs2007}
K.~H. Kim and M.~den Nijs.
\newblock Dynamic screening in a two-species asymmetric exclusion process.
\newblock {\em Phys. Rev. E}, 76:021107, Aug 2007.

\bibitem{Kubo1966}
R.~Kubo.
\newblock The fluctuation-dissipation theorem.
\newblock {\em Rep. Prog. Phys.}, 29:255--284, 1966.

\bibitem{Levine2007}
E.~Levine and T.~Hwa.
\newblock Stochastic fluctuations in metabolic pathways.
\newblock {\em Proc. Natl. Acad. Sci. U.S.A.}, 104(22):9224--9229, May 2007.

\bibitem{Maloney1972}
P.~C. Maloney, E.~R. Kashketl, and T.~H. Wilson.
\newblock A protonmotive force drives atp synthesis in bacteria.
\newblock {\em Proc. Natl. Acad. Sci. U.S.A.}, 71:3896--3900, 1972.

\bibitem{Mangan2003}
S.~Mangan and U.~Alon.
\newblock Structure and function of the feed-forward loop network motif.
\newblock {\em Proc. Natl. Acad. Sci. U.S.A.}, 100(21):11980--11985, Oct 2003.

\bibitem{Oehler1994}
S.~Oehler, M.~Amouyal, P.~Kolkhof, B.~von Wilcken-Bergmann, and B.~Müller-Hill.
\newblock Quality and position of the three lac operators of e. coli define
  efficiency of repression.
\newblock {\em EMBO J}, 13(14):3348--3355, Jul 1994.

\bibitem{Paulsson2004}
J.~Paulsson.
\newblock Summing up the noise in gene networks.
\newblock {\em Nature}, 427(6973):415--418, Jan 2004.

\bibitem{Paulsson2000}
J.~Paulsson, O.~G. Berg, and M.~Ehrenberg.
\newblock Stochastic focusing: fluctuation-enhanced sensitivity of
  intracellular regulation.
\newblock {\em Proc. Natl. Acad. Sci. U.S.A.}, 97(13):7148--7153, Jun 2000.

\bibitem{Pedraza2005}
J.~M. Pedraza and A.~van Oudenaarden.
\newblock Noise propagation in gene networks.
\newblock {\em Science}, 307(5717):1965--1969, Mar 2005.

\bibitem{Rao2003}
C.~V. Rao and A.~P. Arkin.
\newblock Stochastic chemical kinetics and the quasi-steady-state assumption:
  Application to the gillespie algorithm.
\newblock {\em J. Chem. Phys.}, 118:4999, 2003.

\bibitem{Rao2004}
C.~V. Rao, H.~M. Sauro, and A.~P. Arkin.
\newblock Putting the ``control" in metabolic control analysis.
\newblock In {\em 7th International Symposium on Dynamics and Control of
  Process Systems}, 2004.

\bibitem{Scott2007}
M.~Scott, T.~Hwa, and B.~Ingalls.
\newblock Deterministic characterization of stochastic genetic circuits.
\newblock {\em Proc. Natl. Acad. Sci. U.S.A.}, 104(18):7402--7407, May 2007.

\bibitem{Simpson2003}
M.~L. Simpson, C.~D. Cox, and G.~S. Sayler.
\newblock Frequency domain analysis of noise in autoregulated gene circuits.
\newblock {\em Proc. Natl. Acad. Sci. U.S.A.}, 100(8):4551--4556, Apr 2003.

\bibitem{Tanase2006}
S.~T{\u{a}}nase-Nicola, P.~B. Warren, and P.~R. ten Wolde.
\newblock Signal detection, modularity, and the correlation between extrinsic
  and intrinsic noise in biochemical networks.
\newblock {\em Phys. Rev. Lett.}, 97(6):068102, Aug 2006.

\bibitem{Tyson2003}
J.~J. Tyson, K.~C. Chen, and B.~Novak.
\newblock Sniffers, buzzers, toggles and blinkers: dynamics of regulatory and
  signaling pathways in the cell.
\newblock {\em Curr. Opin. Cell. Biol.}, 15(2):221--231, Apr 2003.

\bibitem{Kampen2001}
N.~G. Van~Kampen.
\newblock {\em Stochastic Processes in Physics and Chemistry.}
\newblock North Holland, 2001.

\bibitem{Vilar2002}
J.~M.~G. Vilar, H.~Y. Kueh, N.~Barkai, and S.~Leibler.
\newblock Mechanisms of noise-resistance in genetic oscillators.
\newblock {\em Proc. Natl. Acad. Sci. U.S.A.}, 99(9):5988--5992, Apr 2002.

\bibitem{Warren2006}
P.~B. Warren, S.~T{\u{a}}nase-Nicola, and P.~R. ten Wolde.
\newblock Exact results for noise power spectra in linear biochemical reaction
  networks.
\newblock {\em J. Chem. Phys.}, 125(14):144904, Oct 2006.

\bibitem{Xie2008}
X.~S. Xie, P.~J. Choi, G.-W. Li, N.~K. Lee, and G.~Lia.
\newblock {\em Annu. Rev. Biophys.}, 37:417--444, 2008.

\bibitem{Yu2006}
J.~Yu, J.~Xiao, X.~Ren, K.~Lao, and X.~S. Xie.
\newblock Probing gene expression in live cells, one protein molecule at a
  time.
\newblock {\em Science}, 311(5767):1600--1603, Mar 2006.

\end{thebibliography}
\bibliographystyle{abbrv}

\end{document}